\documentclass[twocolumn]{aastex631}

\makeatletter
\let\frontmatter@title@above=\relax

\usepackage[inline]{enumitem}
\usepackage{amsmath}
\usepackage{multirow}
\usepackage{threeparttable}

\usepackage{xspace}
\newcommand{\ie}{i.e.\/,\xspace}

\newcommand{\eg}{e.g.\/,\xspace}

\newcommand{\grb}{\ensuremath{\mathrm{GRB}}\xspace}
\newcommand{\grbs}{\ensuremath{\mathrm{GRBs}}\xspace}
\newcommand{\lgrb}{\ensuremath{\mathrm{LGRB}}\xspace}

\newcommand{\sn}{SN\xspace}
\newcommand{\sne}{SNe\xspace}

\newcommand{\gamray}{\ensuremath{\gamma}{-ray}\xspace}
\newcommand{\gamrays}{\ensuremath{\gamma}{-rays}\xspace}
\newcommand{\mwl}{MWL\xspace}
\newcommand{\he}{HE\xspace}
\newcommand{\vhe}{VHE\xspace}
\newcommand{\hes}{HEs\xspace}

\newcommand{\ucrs}{UHECRs\xspace}
\newcommand{\snebl}{\sne~Ic-BL\xspace}
\newcommand{\last}{LAST\xspace}
\newcommand{\fov}{\ensuremath{\mathrm{FoV}}\xspace}
\newcommand{\fovs}{\ensuremath{\mathrm{FoVs}}\xspace}
\newcommand{\dgr}{\ensuremath{^{\circ}}\xspace}
\newcommand{\degsqr}{\ensuremath{\mathrm{deg}^{2}}\xspace}
\newcommand{\degsqrinv}{\ensuremath{\mathrm{deg}^{-2}}\xspace}
\newcommand{\yrinv}{\ensuremath{\mathrm{yr}^{-1}}\xspace}
\newcommand{\hrinv}{\ensuremath{\mathrm{hr}^{-1}}\xspace}
\newcommand{\secnd}{\ensuremath{\mathrm{sec}}\xspace}
\newcommand{\secndinv}{\ensuremath{\mathrm{sec}^{-1}}\xspace}
\newcommand{\lumiUnits}{\ensuremath{\mathrm{erg}~\secndinv}\xspace}
\newcommand{\fluxUnits}{\ensuremath{\mathrm{erg}~\mathrm{cm}^{-2}~\secndinv}\xspace}
\newcommand{\snr}{\ensuremath{\mathrm{S/N}}\xspace}
\newcommand{\lumigamma}{\ensuremath{L_{\gamma}}\xspace}
\newcommand{\lumiopt}{\ensuremath{L_{\mathrm{opt}}}\xspace}
\newcommand{\fluxopt}{\ensuremath{F_{\mathrm{opt}}}\xspace}
\newcommand{\lumiratio}{\ensuremath{\mathcal{R}_{\mathrm{o}/\gamma}}\xspace}

\newcommand{\tempProfile}{\ensuremath{\mathcal{T}_{\mathrm{opt}}}\xspace}

\newcommand{\gpcvolinv}{\ensuremath{\mathrm{Gpc}^{-3}}\xspace}
\newcommand{\numDegDot}[2]{\ensuremath{{{#1}.^{\hspace{-3pt}\circ}{{#2}}}}\xspace}

\newcommand{\pixScaleDot}[2]{\ensuremath{{{#1}.\hspace{-2pt}''{{#2}}}}\xspace}
\newcommand{\powA}[2]{\ensuremath{{{#1}\hspace{-1pt}\cdot\hspace{-1pt}10^{{#2}}}}\xspace}
\newcommand{\powB}[1]{\ensuremath{{10^{{#1}}}}\xspace}

\newcommand{\cenkoA}{C09\xspace}
\newcommand{\tensorflow}{\textit{tensorflow}\xspace}

\usepackage{soul}

\begin{document}

\title{Detecting the early optical flashes of gamma-ray bursts with small telescope arrays}

\author[0000-0003-1387-8915]{Iftach Sadeh}
\affiliation{Deutsches Elektronen-Synchrotron DESY, Platanenallee 6, 15738 Zeuthen, Germany}
\email{iftach.sadeh@desy.de}

\begin{abstract}
    We present an observational approach
    for the independent detection of 
    the early optical emission of long
    gamma-ray bursts (\grbs). 
    For this purpose, we 
    explore the potential of the 
    Large Array Survey Telescope (\last).
    This array of small optical telescopes 
    can be used
    to scan a wide region of the sky, and to
    focus on a smaller field of view with increased 
    sensitivity, as needed. The modularity of the array 
    facilitates dynamic scanning of multiple fields,
    by shifting telescope pointing directions
    with high cadence.
    This can
    significantly increase the effective sky coverage
    of a blind survey on short time scales.
    For events associated with gamma-ray counterparts,
    the valuable early time data
    can supplement high-energy observations.
    Regardless of gamma-ray association,
    detections can potentially be used to explore
    various phenomena associated with \grbs,
    such as 
    orphan afterglows;
    dirty fireballs; and
    choked jets.
    We simulate a sample of \grbs and their
    respective optical signals at early times.
    After accounting for dynamic cadence,
    the light curves are given as input to a machine
    learning classifier, used to identify
    astrophysical transients.
    We find that by dedicating half of a \last array to a
    blind search, one would expect
    to independently detect $\text{7--11}$~\grbs per year,
    corresponding to an approximate intrinsic
    event rate of $0.12$~per square degree per year. 
\end{abstract}


\section{Introduction} \label{sec:intro}

Gamma-ray bursts (\grbs) are the most luminous extragalactic
sources in the Universe.
They are caused either by the merging of two compact objects (\eg neutron stars), or
by the core-collapse supernovae (\sne) of massive stripped-envelope 
stars \citep{2015PhR...561....1K, 1993ApJ...405..273W, 1989Natur.340..126E}.
Historically, the two scenarios have
respectively been linked to short- and long-duration \grbs. 
These event-classes respectively  exhibit prompt
emission phases, which are typically
shorter or longer than $\sim2$~\secnd. The prompt
stage is followed by
an extended afterglow phase.
Thousands of \grbs have been detected to date. However, the exact
mechanism of energy dissipation and the associated radiation
processes remain an open question.

During the prompt stage of an event, 
over $10^{51}$~erg of energy
is released as \gamrays. This is thought to be produced by a
relativistic jet that is
launched  by the central engine of the \grb. The jet 
is a collimated outflow of plasma,
having an initial Lorentz factor,
$\Gamma_{\mathrm{init}} \gg 100$. 
The nature of the central engine is unknown, though it
has been hypothesised to be an accreting black hole or a millisecond
magnetar \citep{1994ApJ...430L..93R, 10.1093/mnras/270.3.480}.

At later times, a fraction of the energy of the jet is transformed
to nonthermal radiation as part of the afterglow. As the
\grb outflow sweeps through the external circumburst medium, it is
decelerated by a reverse shock, while a strong
relativistic forward shock propagates outwards.
The latter powers a long-lived emission, 
which is usually attributed to
synchrotron radiation. It is detectable
from radio to X-ray wavelengths on timescales of minutes to
years \citep{1997ApJ...482L..29M}.

The focus of this work is long \grbs.
Long \grbs are predominantly discovered via their prompt emission with
dedicated high-energy (\he) X-ray/\gamray satellites.
In a handful of cases, \grbs have
independently been detected via their
prompt optical emission.
Notable examples include those of
GRB~080319B \citep{2008Natur.455..183R} 
and
GRB~130427A \citep{2014Sci...343...38V}.
However, most events are only observed
in the optical band as part of
multiwavelength (\mwl) follow-up of \he
signals.
Follow-up campaigns involve many
instruments, which are
mainly ground based. These range from the radio
band to very-high-energy (\vhe) gamma-rays. 
The time lag involved
in responding to \grb triggers
from satellites is generally of the order of
tens of \secnd. Follow-up is therefore 
mostly restricted to the afterglow
phase of the emission,
baring extreme
luck, or the availability of
very-wide-field instruments
\citep{2023NatAs...7..724X}).
While very illuminating in itself, the
afterglow in most cases is only indirectly tied to the prompt activity. 
This poses major challenges to our understanding 
of the underlying central engine.

The effective issuance of 
scientific alerts, \eg via the General Circulars Network (GCN),\footnote{The GCN, \url{https://gcn.nasa.gov/circulars}.}
has greatly advanced the field. 
Alerts allow
rapid
response by robotic optical instruments (\eg
\cite{2004SPIE.5489..679S, 2009AJ....137.4100K});
these facilitate \mwl observation
of \grbs in temporal coincidence with their prompt \gamray emission
\citep{2007ApJ...657..925Y}.
Such observations have been used to investigate the
relativistic jet while the central engine was still
active, leading to a variety of
interpretations. In some cases, optical 
flashes were attributed to
internal processes, probing the engine (prompt emission);
in others, they pointed towards
external origins (afterglow emission), \eg a reverse shock
\citep{2013ApJ...772...73K, 2017Natur.547..425T, 2018NatAs...2...69Z, 2021ApJ...908...39B, 2023NatAs...7..724X, 2023NatAs...7..843O}.

Fundamentally, early \mwl data may do more than simply supplement
our observations of the prompt \gamrays. In fact, 
some \grbs do not have an \he counterpart at all.
Such nondetections can come about in different ways, 
the leading explanations being the following: 
\begin{enumerate*}[label=(\roman*)]
    \item unobserved standard \grbs;
    \item orphan afterglows;
    \item dirty fireballs; and
    \item choked jets.
\end{enumerate*}
We discuss each category in the following.

Standard \grbs
may be missed due to
lack of \he sky coverage and
incomplete duty cycles. For example,
\cite{2013ApJ...769..130C} estimate that for events
with fluence, 
${\mathcal{F}_{\gamma} < 6\cdot10^{-7}}$~erg~cm$^{-2}$, 
there is
a~40\% chance of nondetection by the more sensitive
instruments, Fermi-GBM \citep{2009ApJ...702..791M} and
Swift-BAT \citep{2004ApJ...611.1005G}.

Orphan events may
lack \he counterparts due to viewing-angle effects,
given that
\grb jets are highly collimated. To illustrate
the point, the local
\textit{intrinsic} volumetric rate of \grbs is
$79^{+57}_{-33}$~Gpc$^{-3}$~yr$^{-1}$
\citep{2022arXiv220606390G}.
On the other hand,
the corresponding \textit{observed} rate is 
$1.3^{+0.6}_{-0.7}$~Gpc$^{-3}$~yr$^{-1}$
\citep{2010MNRAS.406.1944W}. 
Most events are undetectable in \hes,
as the prompt emission is beamed away from 
our line of sight. 
The \mwl afterglow may nonetheless become visible,
as the jet decelerates and expands sideways.
Prompt low-energy emission may also be observable
for on-axis structured jets that are 
undetected in \gamrays \citep{2010arXiv1012.5101G}. 
In such cases it is
assumed that the relativistic jet is initially observable;
however, it only
contains a narrow core of high Lorentz factor, which 
emits all \gamrays beyond our line of sight.

Dirty fireballs are
a hypothesized class of long
\grbs with low Lorentz factors,
${1 < \Gamma_{\mathrm{init}} \ll 100}$.
They are believed to be
the result of high contents of 
baryonic material entrained in the \grb jet
\citep{2004RvMP...76.1143P}.
These events are still expected to produce
optical signals resembling long \grbs. However,
their \he prompt emission 
might be entirely suppressed \citep{2002MNRAS.332..735H}. Alternatively,
if it exists, the prompt signal would peak below
the nominal sensitivity window of \gamray 
instruments.
Such events go undetected in most cases,
though they might appear as X-ray flashes
\citep{2003AIPC..662..229H, 2005ApJ...629..311S}
or as low-luminosity \grbs
\citep{2017AdAst2017E...5C}.

Choked jets \& shock breakouts fall under 
another postulated scenario,
where a mildly relativistic jet does not manage to
successfully penetrate the stellar envelope
\citep{2017hsn..book..967W}. 
The failed jet dissipates all of its energy 
into the surrounding cocoon, driving it to expand. 
As the cocoon reaches the edge of the star,
a  mild-to-ultrarelativistic 
forward shock emits an X-ray/UV flash
as it breaks out.
This is followed by an extended UV/optical 
signal, arising from the expanding cooling envelope.
These shock breakouts have also been associated with
low-luminosity \grbs. However, as indicated,
the origins and phenomenology
of the prompt emission
are distinct from jetted \grbs \citep{2011ApJ...739L..55B}.

Studying each of the classes of ``nonstandard''
\grbs  can shed light
on fundamental questions in the field.
For instance, while \grbs are disfavoured as
the primary sources of 
ultra-high-energy cosmic rays (\ucrs)
and high-energy cosmic neutrinos, they
may still contribute to these phenomena under
some conditions
\citep{2016PhRvD..93h3003S, 2022MNRAS.511.5823R, 2023ApJ...950...28R}.
Leading uncertainties relate to determining
realistic values for the baryonic loading;
the location of the dissipation region; specifics
of the radiation mechanisms;
the structure of relativistic jets;
the dividing line between successful and failed \grbs;
and the true cosmological rate of events.

These questions are closely tied to the connection between
\grbs and \sne.
Specifically relevant are
\snebl (core-collapse \sne
with broad lines). 
These are stripped-envelope \sne
having systematically high velocities in their optical
spectra (compared to ordinary \sne~Ic at similar epochs).
\snebl have decisively been associated with \grbs 
\citep{2006ARA&A..44..507W}. Determining the comparative
rates between \sne and (low-luminosity) \grbs,
in line with the respective dissipation/radiation models,
will help us understand the physics 
behind massive stellar deaths
\citep{2020ApJ...902...86H, 2023ApJ...953..179C}.

A promising direction to make progress on these
topics is to increase
the number of early time detections of \grbs 
in the optical band. This can
be accomplished by performing a blind search with
a wide-filed-of-view (\fov) telescope,
circumventing the need for satellite-triggers.
If available, detections could be connected after 
the fact to \he and \vhe
counterparts, enabling studies of the
prompt emission spectra \citep{2019A&A...628A..59O}.
A connection could also be made to possible shock breakout
signals via \eg the upcoming ULTRASAT satellite
\citep{2023arXiv230414482S}.
At later times, events would possibly
be associated with \sne via deep optical and radio
follow-up, as well as to \vhe neutrinos.

Multiple groups have attempted to
independently detect \grbs and other short
transients in the optical band
\citep{2018ApJ...854L..13H, 2019MNRAS.484.4507V, 2020MNRAS.491.5852A, 2021AJ....161..135A, 2013ApJ...779...18B}.
For example, the Robotic Optical Transient Search Experiment-III (ROTSE-III)
conducted a search over a \fov of $3.4$~\degsqr with an effective
limiting magnitude, 
$m_{\mathrm{R}}\sim18$, over 60~\secnd
\citep{2005ApJ...631.1032R}. They required
that a transient is detected at least twice, given
a cadence of 30~minutes. This limited the
search to a small area of the sky, and well within the afterglow phase.
Other systems strike a different balance. For instance, Mini-MegaTORTORA
\citep{2017IAUS..324...85K} has
a \fov of 900~\degsqr and
short time resolution.
However, it
is limited to very bright sources,
corresponding to
$m_{\mathrm{V}}\sim13$ over $1$~\secnd
exposures.
Pi of the Sky similarly has a $\sim400$~\degsqr \fov
with a sensitivity,
$m_{\mathrm{V}}\sim12$ over 10~\secnd
\citep{2009AIPC.1133..306S}.

As indicated,
most of these searches have not been
sensitive to rare and faint optical
flashes, shorter than $\sim20$~minutes.
The main challenge has been the design of a system that combines high sensitivity, a wide \fov,
and high cadence.
As a result, only a small sample of untriggered
\grbs have so far been identified through their 
afterglow emission
\citep{2013ApJ...769..130C, 2015ApJ...803L..24C, 2017ApJ...850..149S}. 
The number of events is expected to increase
in the coming years with the advent of
wide-field sky surveys. A prominent example is 
the Zwicky Transient Facility (ZTF),
which has detected 10~afterglows to date
\citep{2022ApJ...938...85H}.

It should be noted
that continuous sky coverage is critical in order
to detect the short prompt emission
of \grbs. 
This is generally
not fulfilled for existing surveys,
optimised for longer, \sn-like
timescales.
For context, the high-cadence programme of ZTF
includes six 30~\secnd visits per night over a wide area of ${\sim2500}$~\degsqr; a fast ZTF
transient is considered
one that fades within a few nights
\citep{2021ApJ...918...63A}.
Excluding lucky coincidences in \fovs,
shorter timescale phenomena 
are nominally studied with
targeted follow-up of known sources
\citep{2020ApJ...896L...2A}. 

It becomes apparent that new experimental
techniques are required in order to
significantly increase our sensitivity
to short optical flashes.
One such approach
is to employ a large number of small
telescopes. An array
of this kind can effectively match
or surpass the capabilities of larger
instruments, but at much lower cost.
This is increasingly being motivated
by the growing availability of
off-the-shelf components, including
back-side-illuminated CMOS detectors 
and fast optical tube assemblies
\citep{2020PASP..132l5004O}.
Examples include
the Ground based Wide Angle Cameras (GWAC), which
is one of the ground facilities of the 
upcoming SVOM mission
\citep{2016arXiv161006892W};
the Gravitational wave Optical Transient 
Observatory (GOTO; \cite{2018SPIE10704E..0CD});
the two Evryscope arrays
\citep{2015PASP..127..234L};
the upcoming Argus array
\citep{2022PASP..134c5003L};
and
the Large Array Survey Telescope (\last;
\cite{2023PASP..135f5001O}).

In this work we propose a blind search for early
optical flashes from \grbs with a small telescope
array, using \last as a benchmark.
If dedicated to \grb science, a \last array
has the potential to
observe a substantial
number of \grbs
during their prompt and afterglow phases.
The paper is organised as follows. 
We begin with an overview of the LAST telescope
array in Sec.~2. In Sec.~3 we present the
simulation parameters of our cosmological \grb 
event sample. This is followed in Sec.~4 with
details on the simulation of the early optical emission
for each event. In Sec.~5 we discuss the backgrounds 
relevant for a blind optical survey on short
time scales. In
Sec.~6 we describe the procedure for conducting
the survey, and for using it to detect \grbs. 
We conclude with a summary and
discussion of the results in Sec.~7.

\section{The \last telescope array}
\label{sec:last_instrument}

\begin{figure}[tp]
    \centering\includegraphics[trim=0mm 0mm 0mm 180mm,clip,width=.47\textwidth]{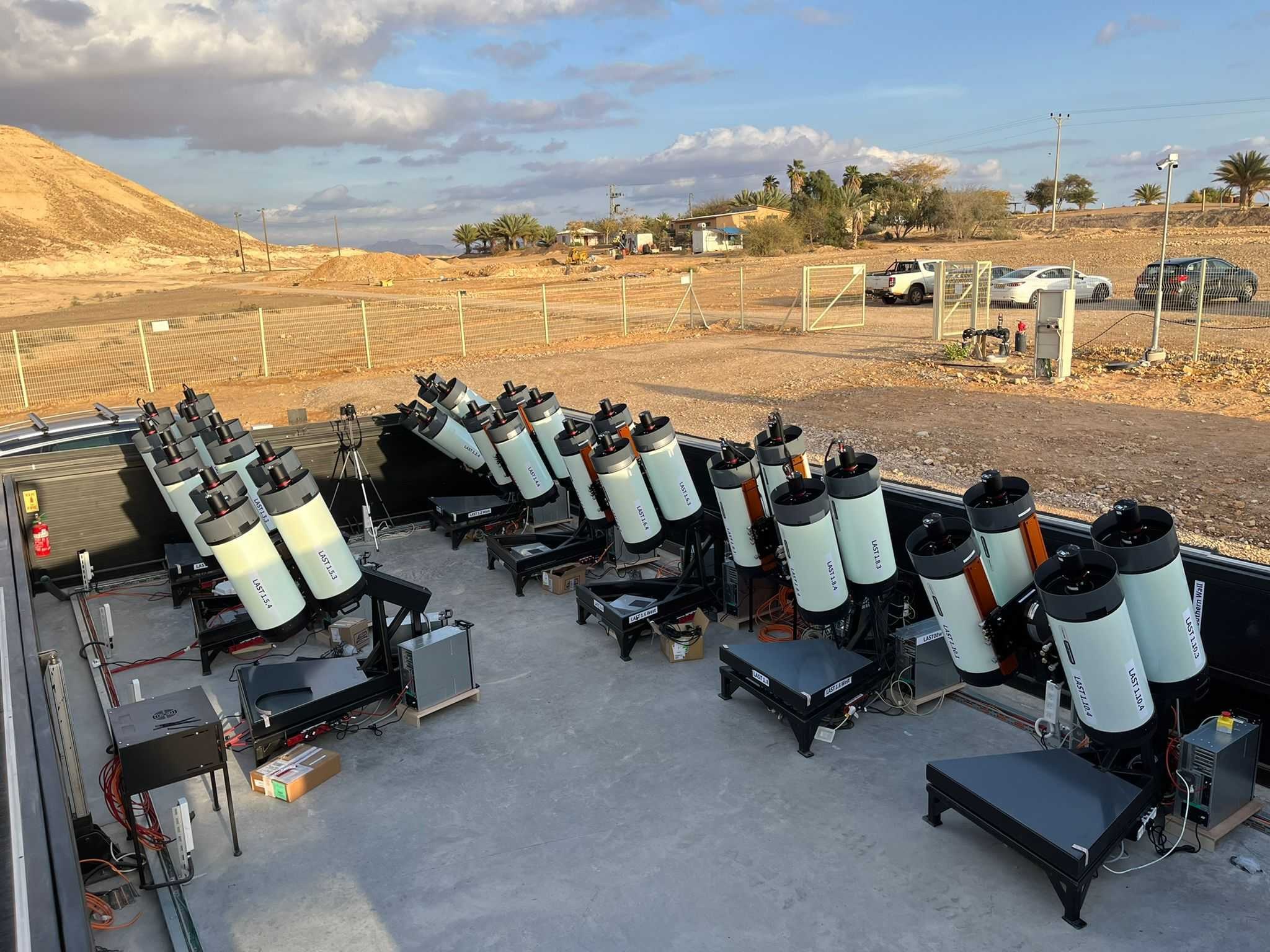}
    \caption{
        The first node of \last at the Weizmann Astrophysical Observatory, with the enclosure fully opened. The array here includes
        32 of the planned 48~telescopes
        already installed. (See \cite{2023PASP..135f5001O}.)
    \label{fig:last_node}}
\end{figure}

\last is an array of small optical telescopes.
The first node of \last is currently being instrumented
at the Weizmann Astrophysical Observatory in the south 
of Israel, as shown
in Fig.~\ref{fig:last_node}, and
described in \cite{2023PASP..135f5001O}. 
Additional facilities are
being planned, pending funding.

A single node of \last nominally comprises 
48 telescopes, installed on 12 independent mounts,
which are housed inside a rolling-roof enclosure.
LAST uses
Xerxes equatorial mounts.
These employ a pair of direct-drive 
motors, which can produce torques of 60 N-m (hour angle), and 33 N-m (decl.).
\last telescopes are 27.9~cm $\mathrm{F/2.2}$ 
Rowe-Ackermann Schmidt Astrographs (RASA) from Celestron.
Each has a \fov, $\numDegDot{3}{3} \times \numDegDot{2}{2} \approx 7.4$~\degsqr,  when coupled to 
full-frame (${36 \times 24}$~mm) detectors.
The cameras are of type QHY600M, with back-illuminated thermoelectrically cooled Sony IMX-455 CMOS sensors.

The image quality, including seeing effects, results
in a resolution of
\pixScaleDot{2}{2}--\pixScaleDot{2}{8}
for sources near the centre of the field.
The typical $5\sigma$ limiting magnitude 
of a single telescope without any filter
is $m_{\mathrm{inc}} = 19.6$ over $20$~\secnd exposures.
This bandpass resembles the Gaia 
$G_{\mathrm{bp}}$~band
\citep{2016A&A...595A...1G}, 
as measured by \cite{2023PASP..135f5001O}.

The modular nature of \last provides flexibility
for conducting wide and deep surveys. Using the
so-called ``wide'' observing mode, each telescope
points at an independent field across the sky.
This provides a simultaneous ${\sim355}$~\degsqr \fov
($0.8\%$ of the celestial sphere) at the nominal
depth of $19.6$~mag.
Alternatively, several or all telescopes may 
be pointed at the
same coordinates using the
``narrow'' observing mode. This substantially
improves the sensitivity,
comparable to a $1.9$~m telescope. 
For instance, coadding
$20$ of the telescopes achieves a limiting magnitude,
$\mathrm{m}_{\mathrm{inc}} = 21$ (20~\secnd; ${7.4}$~\degsqr).

While LAST mounts can slew
between different sky regions, individual 
telescope alignments are fixed.
Manual intervention is needed
in order to transition elements
of the array between
wide and narrow modes.
This involves
modifying the physical placement of
telescopes on their mounts, followed
by a software-driven alignment and
calibration procedure.
It is therefore impractical to
change the observing mode of a 
particular mount 
during a night of automated operations. 
Correspondingly, we assume in the following
that half of a \last node is prearranged
with the wide observation mode, and the other
half uses the narrow mode.

The \last cameras use a rolling shutter, which
enables continuous readout
with negligible dead time between exposures. 
It takes $0.7$~\secnd to read images into memory, and up to an additional $1.5$~\secnd to
write them to disk.
While in principle, this allows one to 
produce images with $\sim1$~\secnd resolution,
it is advisable to choose slightly longer
exposures. The primary reason is that 
the transition from read-noise noise to 
background-dominated noise takes place at 
exposure times of $\sim5$~\secnd. 
The mounts can move very quickly. For safety reasons, slewing is currently limited 
to speeds of up to $12\dgr~\secndinv$.
It takes up to $\sim2$~\secnd to slew to an 
adjacent \fov, stabilise, and begin tracking.

A rich scientific agenda is planned for 
the first \last node, as detailed in
\cite{2023PASP..135h5002B}. The planned survey
will be dedicated to searches for 
gravitational-wave (GW) electromagnetic counterparts;
the study of planetary systems around white dwarfs;
the search for near-Earth objects;
\sne science;
and
follow-up of \he neutrinos,
to name just a few cases.
The nominal observation pattern will involve high-
and low-cadence surveys.
For illustration, the high-cadence option
will include eight visits per night per pointing;
each visit will comprise
$20$~consecutive exposures of $20$~\secnd, which
will be coadded.
A small fraction of the time will be devoted to
target of opportunity (ToO) observations,
such as follow-up of \grb alerts.

In the following, we propose an alternative observation
strategy, which is optimised for serendipitous
discovery of optical \grb counterparts. This can be used part of the time
and/or with a subset of telescopes from the first
\last node. The same could also be the 
focus of another dedicated instrument.
In order to inform the design of
our survey, we simulate the expected 
\grb event rate, as described in the following.

\section{Cosmological event rate} 
\label{sec:grb_event_rate}

The number of long \grbs per unit time at 
redshift, $z \sim z + dz$,
with luminosity, 
$\lumigamma \sim \lumigamma + d\lumigamma$, 
is given by
\begin{equation}
    \frac{dN}{dz\; d\lumigamma\; dt} =
        \frac{
            R_{\lgrb}(z)
        }{
            1 + z
        } 
        \frac{dV(z)}{dz}
        \Phi(\lumigamma)
        \;.
\label{eq:grb_number_density}
\end{equation}
We define $R_{\lgrb}$ as the volumetric event rate of
long \grbs per unit time; 
the factor, $(1 + z)^{-1}$, accounts 
for cosmological time dilation;
$\Phi$ is the luminosity function of long \grbs;
and $dV / dz$
is the comoving volume element at redshift, $z$.
We assume a flat, $\mathrm{\Lambda CDM}$ universe, where
${H_{0} = 67.7\;\mathrm{km}\;\secndinv\;\mathrm{Mpc}^{-1}}$,
$\Omega_{\mathrm{M}} = 0.31$, and
$\Omega_{\mathrm{\Lambda}} = 0.69$
\citep{2020A&A...641A...6P}.

Given the connection of long \grbs to \sne,
it is generally accepted
that $R_{\lgrb}$
follows the cosmic star formation rate, 
$R_{\mathrm{SFR}}$, \citep{2008ApJ...673L.119K}
where
\begin{equation}
    R_{\lgrb} = 
        \rho{_0} \cdot 
        R_{\mathrm{SFR}} \cdot 
        f_{\mathrm{ev}} \cdot 
        \Theta_{M} \;.
\label{eq:grb_event_rate}
\end{equation}
Here $\rho{_0}$ is the local \grb event rate,
and
\begin{equation}
    R_{\mathrm{SFR}} \propto  
        \begin{cases}
            (1 + z)^{3.44},  
                & z \leq z_{\mathrm{peak}} \\
            (1 + z_{\mathrm{peak}})^{3.44},
                & z > z_{\mathrm{peak}} \;, \\
        \end{cases}
\label{eq:sfr_rate}
\end{equation}
given $z_{\mathrm{peak}} = 1$
\citep{2006ApJ...651..142H, 2008ApJ...673L.119K}.
We account
for a possible evolution effect in excess 
of the star formation rate,
\begin{equation}
    f_{\mathrm{ev}} = (1 + z)^{\delta}
\label{eq:grb_redshift_evolution}
\end{equation}
with $\delta = 0.4$
\citep{2010MNRAS.406..558Q}. Finally,
$\Theta_{M}$ is the fractional galaxy mass density
as a function of metallicity, as defined in Eq.~5 of
\cite{2006ApJ...638L..63L}.

We model the luminosity function as 
a two-component broken power-law
\begin{equation}
    \Phi(\lumigamma) \propto 
        \begin{cases}
          \left(\frac{\lumigamma}{L_{c}}\right)^{-a}, 
            & \lumigamma \le L_{c} \\
          \left(\frac{\lumigamma}{L_{c}}\right)^{-b}, 
            & \lumigamma > L_{c} \;, \\
        \end{cases}
\label{eq:grb_luminosity_func}
\end{equation}
where 
$L_{c} = \powA{2.5 ^{+6.8} _{-2.1}}{52}$ 
represents the break luminosity,
$a = 1.56 ^{+0.11} _{-0.42}$, and
$b = 2.31 ^{+0.35} _{-0.31}$
\citep{2012ApJ...749...68S, 2015MNRAS.447.1911P}.

We consider here the nominal
luminosity range for standard \grbs
(excluding low-luminosity bursts),
which is
$\log_{10}(\lumigamma \; [\lumiUnits]) \in [50, 54]$.
The luminosity function is normalised, such that
integrating the number density over the local
volume recovers the observed event rate,
$\rho{_0} = 1.3$~\gpcvolinv~\yrinv 
\citep{2010MNRAS.406.1944W}.
In the following, we constrain ourselves to
redshifts, $z < 6$. This is motivated
by our interest in
the optical emission of \grbs, which is 
highly suppressed by
Lyman~$\alpha$ absorption at higher redshift
\citep{2000ApJ...536....1L}.

\section{Optical GRB signatures} 
\label{sec:optical_signatures}

\subsection{Luminosity calibration}

The objective of this study
is to understand the rate of possible \grb detections
with \last. 
The \he emission of events should therefore
be connected to observable optical signals.
In principle, one could model the early optical emission
of \grbs for each and every event
as a low-energy extension of the
prompt phase, or as part of the afterglow
\citep{2019A&A...628A..59O, 2013ApJ...772...73K}.
However, we are currently only interested in the average
properties of the entire sample.
We can therefore take a simpler, data-driven approach,
as described in the following.

The relation between the optical luminosity, \lumiopt,
and the corresponding flux, \fluxopt, is
\begin{equation}
    \lumiopt = 
        4\pi \; D_{L}^{2}(z) \;
        \fluxopt \; K(z)
        \;.
\label{eq:optical_lumi_to_flux}
\end{equation}
The luminosity distance at a given 
redshift is denoted by $D_{L}$; 
the \textit{k}-correction, $K$, accounts
for the shift in frequency between the
observed and rest frames of reference
\citep{2002astro.ph.10394H}.
We can derive the flux from the corresponding
\gamray luminosity as
\begin{equation}
    \fluxopt = 
        \frac{
            \lumigamma \; \lumiratio
        }{
            4\pi \; D_{L}^{2} \; K
        }
        \;,
\label{eq:gamma_lumi_to_flux}
\end{equation}
where $\lumiratio = \lumiopt / \lumigamma$ denotes
the ratio of optical to \gamray luminosities.

We assume the following model for the 
spectral flux density in the optical band,
given frequency, $\nu$, and time, $t$:
\begin{equation}
    f_{\mathrm{opt}}(\nu,t) \propto
        \nu^{-\beta_{0}} 
        \cdot \tempProfile(t)
        \;.
\label{eq:optical_spec_model}
\end{equation}
Here \tempProfile is the temporal profile;
the spectral index, $\beta_{0}$, is drawn
from a normal distribution with parameters,
$0.79 \pm 0.03$
\citep{2020ApJ...905L..26D}.
The corresponding \textit{k}-correction is
\begin{equation}
    K =
        \frac
        {1}
        {
            (1 + z)^{1 - \beta_{0}}
        }
        \;.
\label{eq:optical_k_correction}
\end{equation}

Our point of reference for the analysis is
the study of \cite{2009ApJ...693.1484C}, 
hereafter denoted by \cenkoA. \cenkoA compiled  
a sample of Swift-detected
early afterglows with the 
Palomar 60~inch robotic telescope (P60).
They derived the optical luminosity
in the $R_{\mathrm{C}}$ band
for a common rest-frame time with respect to
the beginning of the burst, $t = \powB{3}$~\secnd.
The observations could be described by
a log-normal distribution with mean,
$\log_{10}(L_{\mathrm{R}}\;[\lumiUnits]) = 46.68$,
and a standard deviation of $1.04$~dex.
In the following, we assume that such a luminosity
distribution is representative 
at this epoch; we use it
to derive the normalisation scale for
Eq.~\ref{eq:optical_spec_model}, as discussed in
Sec.~\ref{sec:survey_light_ccurve_simulation}. 
It remains to define \tempProfile,
in order to
extrapolate the emission to earlier times,
as described next.

\subsection{Temporal profiles} 

The phenomenology of optical \grb light curves is very rich. For instance, 
\cite{2008MNRAS.387..497P,2011MNRAS.414.3537P}
identify three broad classes of the early
emission, which they denote as 
\textit{fast/slow rises};
\textit{plateaus}; and
\textit{decays}. 
The \textit{raising} and \textit{plateauing} 
light curves may \eg be
attributed to geometric viewing-angle effects,
or to shocks generated as the ejecta are 
decelerated by the external medium as part
of the afterglow. 
Such profiles generally exhibit structured
features on time scales of up to \powB{3}~\secnd.
At early times, they may also
be directly related to the prompt 
activity of the central engine of the \grb.
The class of \textit{decays}
corresponds to
about 50\% of events \citep{2023Univ....9..113O}.
These \grbs are characterised by a simple
power-law decay of the optical flux with time. 
They presumably include a very
fast-rising emission
phase, occurring
before the start of observations. 
(For instance, such a fast rise
was observed for GRB~160625B by \cite{2017Natur.547..425T}.)

Our focus for the current study is on very
early optical detection (within tens of~\secnd), 
which would potentially coincide
with the prompt \he emission. We therefore
model the temporal profile of \grbs based
exclusively on
the \textit{decay} class.
We assume a combination of
smoothly connected segments,
\begin{equation}
    \tempProfile(t) = 
        \begin{cases}
          a_{t} + b_{t} \cdot t \;, 
            & t \le \tau_{0} \\
          c_{t} \;, 
            & \tau_{0} < t \le \tau_{1} \\
          c_{t} \cdot (1 + t - \tau_{1})^{-\alpha_{t}} \;, 
            & \tau_{1} < t \le \tau_{2} \\
          d_{t} \cdot (1 + t - \tau_{1})^{-\beta_{t}} \;, 
            & t > \tau_{2} \;. \\
        \end{cases}
\label{eq:temporal_profile}
\end{equation}
These correspond to the following:
\begin{enumerate}[label=(\roman*)]
    \item a fast-rising pulse with
    duration, $\tau_{0}$;
    the constants, $a_{t}$ and $b_{t}$,
    determine a linear increase in flux from a relative
    value of 10--50\% with respect to the peak;
    \item a short plateau with
    duration, 
    $(\tau_{1} - \tau_{0})$,
    having constant peak flux, $c_{t}$;
    \item an initial power-law decay phase
    having a temporal index, 
    $\alpha_{t}$;
    the duration of this phase
    relates to the typical prompt 
    duration expected for long \grbs, denoted by
    $\tau_{2}$; and
    \item a long-lasting second power-law decay 
    phase, which extends into the afterglow; this phase
    is characterised by a temporal index, 
    $\beta_{t}$; it is
    smoothly connected to the early decay
    via the constant, $d_{t}$.
\end{enumerate}

\begin{figure}[tp]
    \centering
    \includegraphics[trim=0mm 0mm 3mm 0mm,clip,width=.48\textwidth]{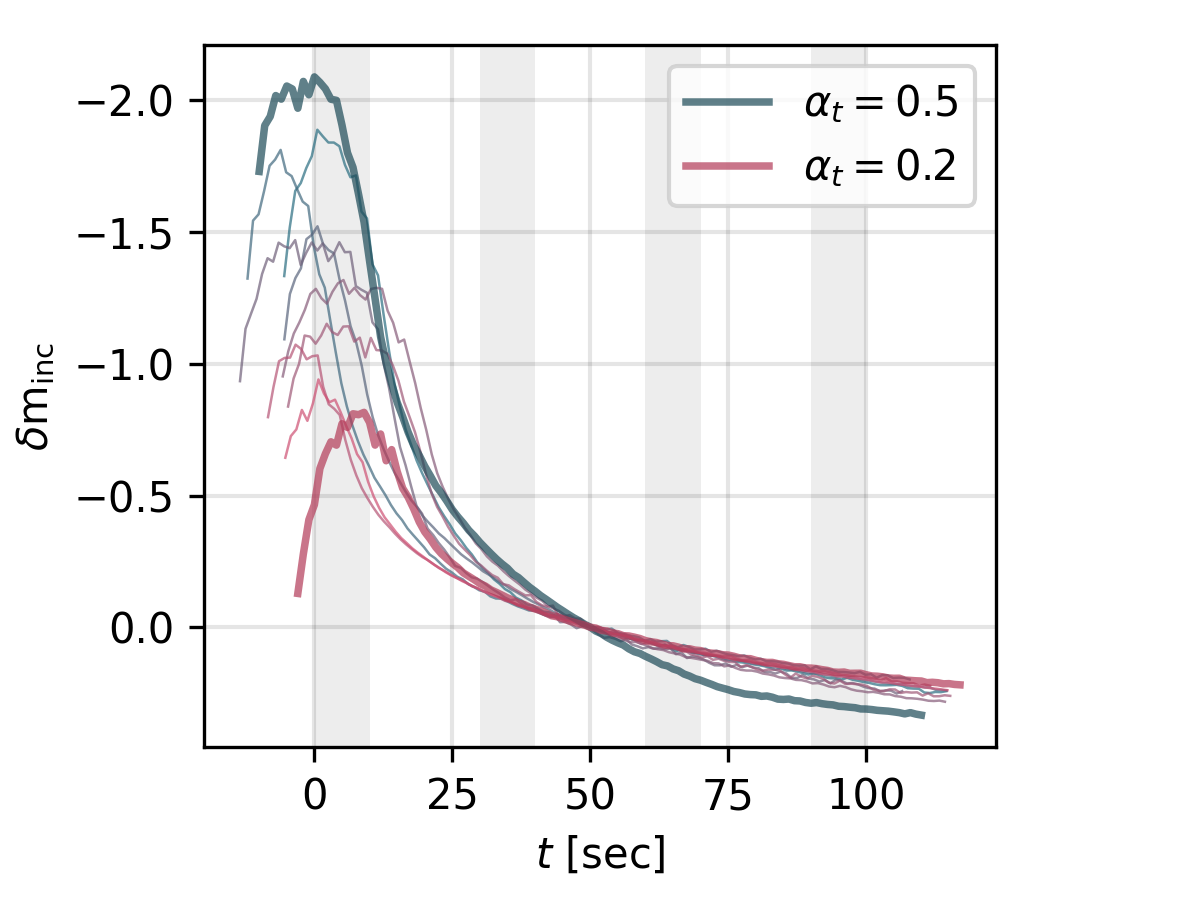}
    \caption{
            Examples of \grb temporal profiles, \tempProfile,
            represented as the change in (unfiltered) 
            \last magnitude, 
            $\delta m_{\mathrm{inc}}$, as a function of time,
            $t$, with respect to the presumed beginning of
            observations. The different curves illustrate
            various realisations, sampled from
            the parameter space
            detailed in Table.~\ref{table:survey_summary}. 
            Two curves
            are highlighted, corresponding to the minimal
            and maximal values of the initial
            temporal decay index, $\alpha_{t}$, as indicated.
            The vertical gray bands illustrate
            $10$~\secnd exposure
            intervals, interleaved with $20$~\secnd
            gaps. This corresponds to our initial observation
            pattern, as discussed in Sec.~\ref{sec:survey_cadence}.
            For visual clarity, the presented profiles 
            are normalised to a common
            point of reference, such that $\delta m_{\mathrm{inc}} = 0$
            for $t = 50$~\secnd. The actual reference time
            used in this study is \powB{3}~\secnd, where
            the relative normalisation is determined
            by the optical flux, as detailed in the text.
    \label{fig:light_curves}}
\end{figure}

We generate different realisations of temporal 
profiles per \grb, based on randomised values of the
parameters, as listed
in Table~\ref{table:survey_summary}.
In addition, we add noise on the level of a few percent
to each phase of the profile. 
A representative sample
of light curves is shown in Fig.~\ref{fig:light_curves}.
As indicated, 
the choice of temporal decay indices is impactful. 
Choosing a steep spectrum (\ie a high
value of $\alpha_{t}$) results in significant relative enhancement
of the early time flux. We correspondingly
choose a range of values that limits the variance
of our sample in brightness.
On average, 
peak magnitudes are about $2$~mag brighter
than those at $t = \powB{2}$~\secnd, and 
about $3$~mag brighter 
than those at $\powB{3}$~\secnd.

\section{Backgrounds to a blind search} 

The main sources of background for \grb optical
flares on second-time scales are 
\begin{enumerate*}[label=(\roman*)]
    \item cosmic rays;
    \item geosynchronous and graveyard-orbit satellites;
    and
    \item stellar flares, mostly from M-dwarfs.
\end{enumerate*}

Cosmic rays lose energy when hitting a detector,
which may cause a bright spot to appear in one
or two consecutive images.
In general, taking multiple exposure is an effective way
to reject this background. Additionally, cosmic
rays may be  identified in some cases in a single 
image by their shape, which is sharper than that
of astrophysical point sources
\citep{2001PASP..113.1420V}.
The rate of cosmic ray artefacts in a
typical image is about
$0.07$~\degsqr~\secndinv.

Satellites may also mimic transient
astrophysical sources, exhibiting flashes with 
durations of $0.2$~\secnd and 
a brightness of 9--11~mag
\citep{2021MNRAS.505.2477N}.
Low-Earth-orbit satellites move at velocities
of hundreds of arcseconds per second and would appear
as streaks in an image. 
On the other hand, satellites at high orbits may seem
motionless. 
They can manifest themselves as single
flashes. They can also appear
as repeating flashes with similar
magnitude, mapping a straight
line across multiple images.
The rate of occurrence of such flashes,
which depends on declination, 
is $\sim1.5$~\degsqrinv~\hrinv for individual flares,
and $\sim0.2$~\degsqrinv~\hrinv for repeaters.

Stellar flares are triggered by magnetic reconnection 
in the corona \citep{2011LRSP....8....6S}.
Flares in the visible band
may last from minutes to hours. They
are modelled by a blackbody spectrum, and tend
to follow fast-rise and exponential-decay
profiles.
Small flares occur much more 
frequently than large ones.
Based on dimensional arguments and observational
data, the rise-time of such flares
scales with their energy
as $E_{\mathrm{flare}} \propto t_{\mathrm{rise}}^{3}$
\citep{2022PASJ...74.1069A}.
So-called ``superflares'', which include an increase
in brightens larger than $3$~mag, are observed
several times a year by very wide-field
surveys \citep{2019ApJ...881....9H}.

Both cosmic ray and satellite backgrounds to \grbs may
effectively be identified and rejected individually,
or by taking multiple short consecutive images.
Stellar flares require further attention. 
The lion's share of
events may be rejected by cross-matching their
location with known stars. For instance,
the Gaia catalogue of stars is complete up to
at least $m_{\mathrm{G}}\sim20$
in noncrowded sky regions
\citep{2020MNRAS.497.4246B, 2023A&A...674A...1G}.
While fainter sources may only become visible
as they flare, their  
temporal profiles are expected to be distinct
from those of \grbs.

A direct consequence of the above,
is that \grbs can not confidently be identified 
based on a single bright flash. 
Rather, it is necessary to obtain 
fine-grained temporal light curves, 
as well as long-term follow-up observations.
In the following, we show how an instrument
such as \last may be used to obtain these early data.

\section{Survey simulation} 
\label{sec:survey_simulation}

\subsection{Light-curve simulation} 
\label{sec:survey_light_ccurve_simulation}

Observationally, a wide diversity of 
optical to \gamray luminosity ratios
has been observed, spanning several orders of magnitude
\citep{2013ApJ...772...73K}.
Since we do not attempt to model the 
spectral energy distribution for
individual \grbs, the value of \lumiratio
is not determined from first principles. 
Instead, we assign a flux ratio probabilistically,
under the assumption that the
\lumiopt distribution of \cenkoA is universal.

We do not assume a specific correlation between
the brightness of events in the two bands,
in accordance with observations
\citep{2007ApJ...669.1107Y}.
Rather, we first generate independent collections
of \gamray luminosity distributions; 
we apply a
given value of \lumiratio to each; and finally, we fit the 
combination of samples to the reference distribution.
We found that our initial simulation slightly overestimated the rate of bright events.

This is likely due to a small mismatch in
the redshift distributions between our simulation
and the \cenkoA dataset. 
We corrected this 
and simulate slightly dimmer optical emission,
by shifting the reference distribution by
$-0.5$~dex.
(The compatibility of the final simulated
brightness distribution is
verified in Fig.~\ref{fig:cumulative_dist_t_1000} below.)
\begin{figure}[tp]
    \centering
    \includegraphics[trim=0mm 0mm 3mm 0mm,clip,width=.48\textwidth]{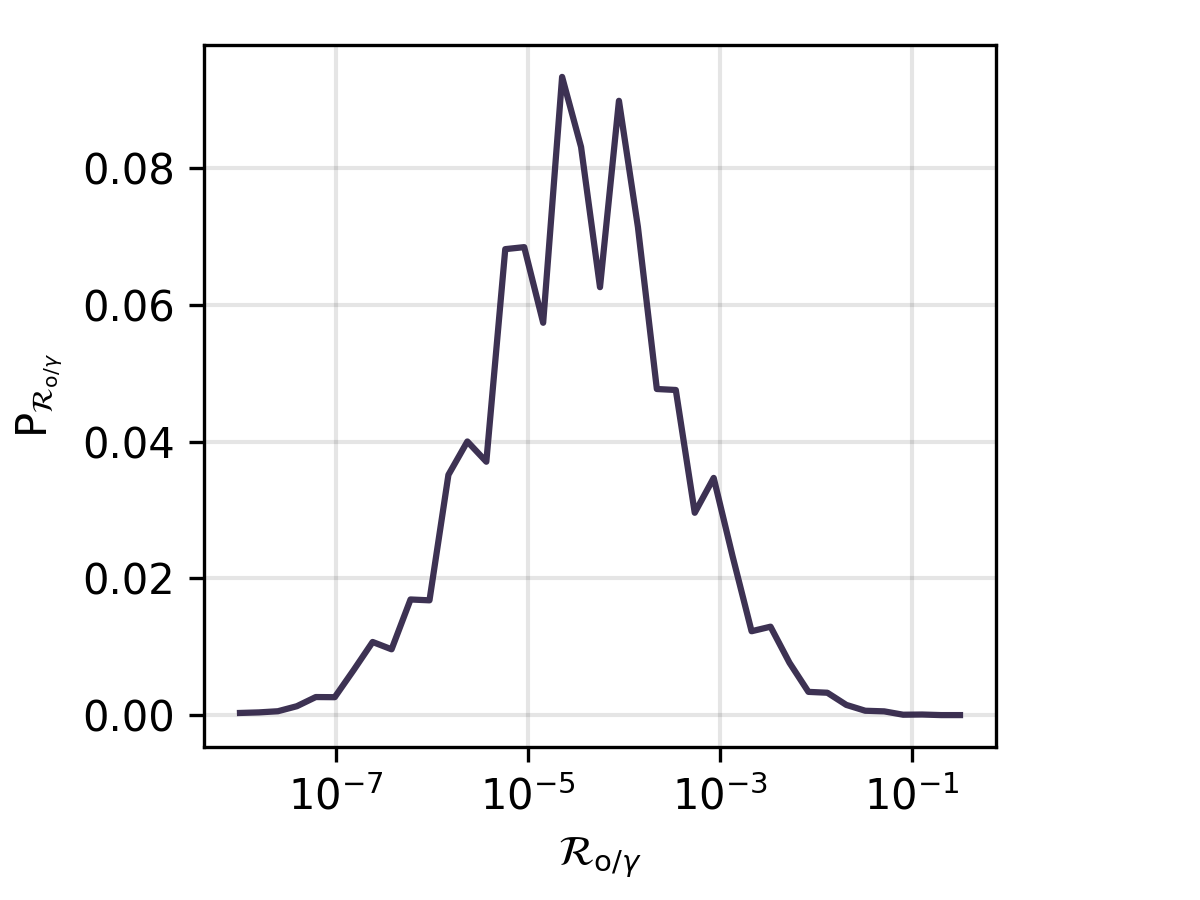}
    \caption{
        Distribution of the derived
        ratio of optical to \gamray luminosities
        at \powB{3}~\secnd, 
        where $\mathrm{P}_{\lumiratio}$ denotes
        the probability density function of \lumiratio.
    \label{fig:gamma_opt_lumi_weights}}
\end{figure}
The results for the luminosity ratio are shown in Fig.~\ref{fig:gamma_opt_lumi_weights},
where the distribution peaks for values,
$\lumiratio \sim \text{\powB{-5} -- \powB{-4}}$.

We continue and
generate a uniform distribution of \grbs in a 
fine-grained grid in luminosity, \lumigamma, 
and redshift, $z$. Events are
weighted according to their relative number
density, per Eqs.~\ref{eq:grb_number_density}--\ref{eq:grb_luminosity_func}.
This process is repeated for different optical to
\gamray luminosity
ratios; each \grb is reweighted by the probability 
distribution for the corresponding \lumiratio.

A simulated \grb is defined by the set of 
redshift, luminosity, \lumiratio,
$\beta_{0}$, and a particular realisation of \tempProfile.
We derive the corresponding
flux normalisation for Eq.~\ref{eq:optical_spec_model}
by integrating the spectrum over the $\mathrm{R}_{\mathrm{C}}$ band. This
allows us to derive the respective flux and
magnitude, $\mathrm{m}_{\mathrm{inc}}$,
over the inclusive (unfiltered) bandpass of \last.
We dim the observed emission, assuming
an extinction value, $A_{\mathrm{V}}$, drawn from a
reference distribution
($\log_{10}A_{\mathrm{V}} = -0.63 \pm 0.42$;
\cite{2013ApJ...774..132W}).
Compared to the prompt phase,
our reference luminosity represents a relatively late epoch
of $\powB{3}$~\secnd. We extrapolate the flux
for the early emission according to
\tempProfile.

\subsection{Survey cadence} 
\label{sec:survey_cadence}

The potential of a blind survey to detect
the  optical 
flares of \grbs is directly proportional to
the accessible search volume, and to
the sensitivity to faint signals.
Different arrangements of telescopes
may be used to balance between
these two factors.
As discussed \eg by \cite{2003AJ....125.2740N},
it is possible to optimise the observing pattern,
based on a set of simple assumptions.
For instance, these include the
spectro-temporal properties 
of the emission; the background
conditions; the limiting magnitude
of telescopes;
and their slewing speed.
One may subsequently 
maximise the number of putative
detections for later follow-up.

In the current study, we illustrate
a survey strategy that balances between
source discovery and temporal sampling of the
respective light-curve. 
We initially scan a large area of
the sky intermittently. This is done using
half of a \last array, arranged in the
wide observing mode. 
Given the detection of candidate
events, we focus the other half of the
array, arranged in the narrow mode,
on the relevant \fov. This increases
the sensitivity while observing the
rapidly decaying emission of the source.
Probing the light-curve several times
has the advantage of reducing the
number of fake positive detections, 
as discussed below.

\begin{figure*}[tp]
    \begin{center}
        \includegraphics[trim=0mm 0mm 0mm 0mm,clip,width=0.85\textwidth]{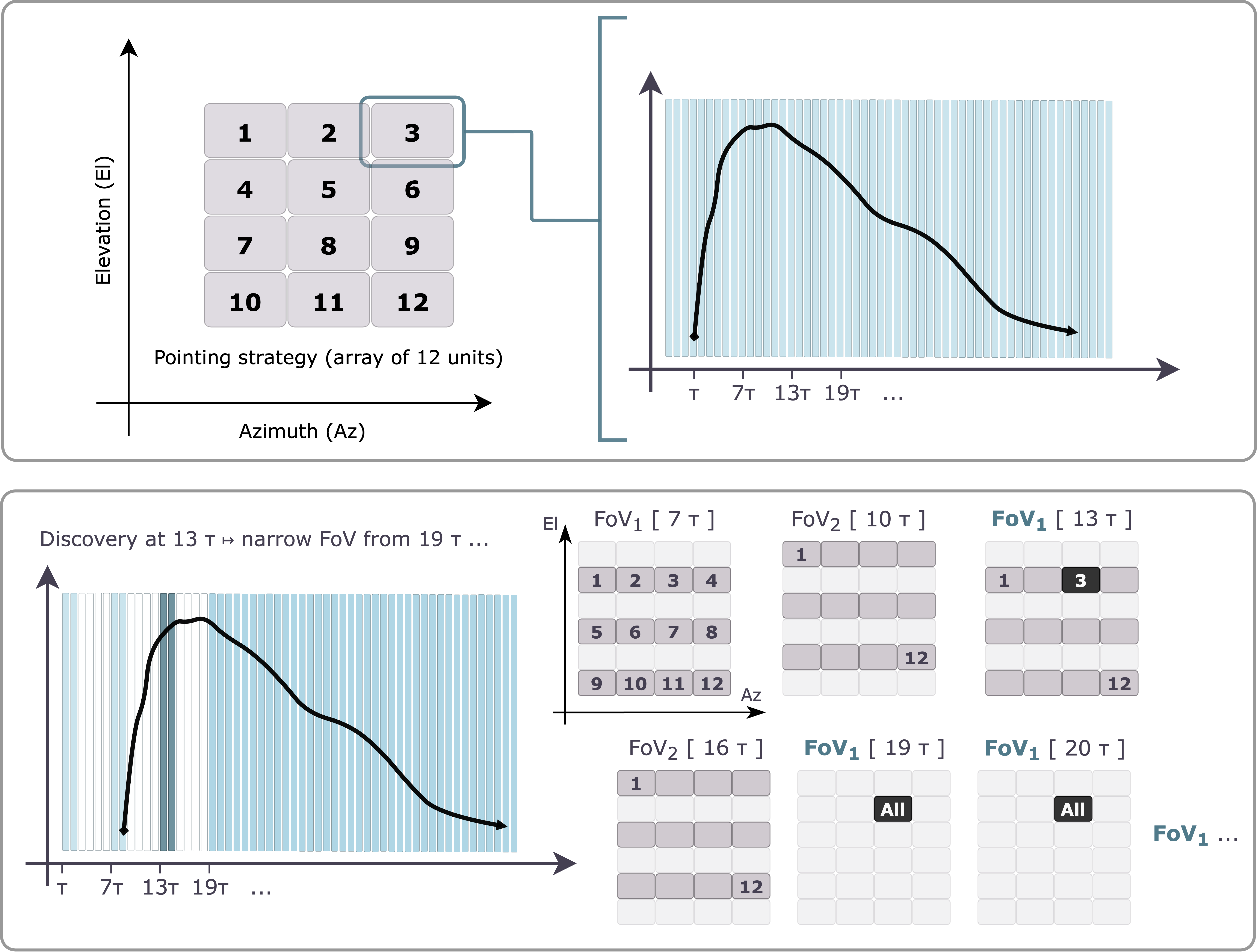}
        \caption{
            Schematic representation of two blind survey strategies. For visual clarity, we
            illustrate here a \last array consisting
            of only 12~telescopes. 
            The top panel illustrates
            a ``simple'' wide observing layout, where the different
            telescopes are deployed in a continuous 
            ${3 \times 4}$ grid in azimuth and elevation. 
            Highlighting the \fov of one of the telescopes (denoted 
            here as ``$3$''), the light-curve of a putative
            transient source is continuously sampled in steps (exposures)
            of duration, $\tau$.
            The bottom panel illustrates
            a ``dynamic'' observing strategy, where each telescope
            transitions between two \fovs, denoted collectively as
            $\fov_{1}$ and $\fov_{2}$. This corresponds to an initial sparse
            observing pattern for each \fov.
            At some point in time, denoted here as $13\tau$,
            a flare candidate is detected. Following a short interval for the candidate to be
            identified, multiple telescopes in the
            narrow alignment mode
            converge on the relevant field. 
            The light
            curve is further sampled without gaps.
        \label{fig:last_survey_cadence}}
    \end{center}
\end{figure*}

We design the observing strategy 
with the objective of effectively
doubling the baseline sky coverage of \last. 
This may be accomplished by taking 
advantage of the rapid slewing capabilities of
the instrument, as well as the option to 
deploy subsets of telescopes
in wide and narrow 
observing modes.
Using a simple wide layout of telescopes, it is possible
to continuously tile a region of the sky. 
This is illustrated in the
top panel of Fig.~\ref{fig:last_survey_cadence}. 
In this case, each \fov is
continuously being observed using short 
exposures of $\tau_{\mathrm{obs}}$~\secnd.

Alternatively, it is possible to deploy telescopes
with wider margins, as \eg shown in the bottom panel of
the figure. Here we illustrate the following observing 
pattern:
\begin{enumerate*}[label=(\roman*)]
    \item take a pair
        of exposures 
        ($2 \times \tau_{\mathrm{obs}}$) 
        of a particular field,
        designated as $\fov_{1}$;
    \item slew to a different field, 
    $\fov_{2}$; this incurs a gap in observations of
    $\tau_{\mathrm{slew}}$~\secnd;
    \item take a pair
        of exposures, pointing at $\fov_{2}$;
    \item slew telescopes back to their original coordinates.
\end{enumerate*}
This pattern constitutes a single observing cycle, which is 
continuously repeated until a source-candidate is detected. 

For the
current study, we assume that
the \last array comprises $40$~telescopes, $20$~of
which are aligned in the wide
observing mode and participate in the blind search.
The other set of $20$~telescopes is aligned in
the narrow mode.
After a potential flare is identified,
the subarray of narrow/convergent 
telescopes is repointed onto
the respective \fov. 
This substantially increases the 
sensitivity to the rapidly fading emission of the
putative \grb.

We choose
$\tau_{\mathrm{obs}} = 5$~\secnd and
$\tau_{\mathrm{slew}} = 5$~\secnd,
which constitutes a ${\text{1:2}}$~cadence pattern. 
Correspondingly,
we take exposures of $10$~\secnd 
during the blind search stage
(coadding $2 \times \tau_{\mathrm{obs}}$ exposures),
which are
separated by $20$~\secnd gaps.
These include slewing intervals
and observations of alternative fields.
For example, a possible light-curve of exposures
might correspond to 
$t \in$~[0--10, 30--40, 60--70, 70--80, 80--90, ...]~\secnd;
this assumes a transition to narrow observing 
at $t = 60$~\secnd.
We explicitly choose a pair of $5$~\secnd exposures,
rather than a single $10$~\secnd one, which
mitigates cosmic ray and satellite 
backgrounds.

The dynamic observing approach increases
the number of observable events, by
compromising on detection of
a short segment of the initial light-curve.
The success of such a strategy depends on development
of efficient analysis tools, which enable fast
coaddition of images and
identification of flares. 
We note that the 
requirement on the precision and false-positive
rate of this initial filter is not stringent,
given that artefacts would quickly be identified after
transition to the narrow observing mode.

\subsection{Flare classification} 
\begin{figure*}[tp]
    \centering
    \includegraphics[trim=0mm 0mm 3mm 0mm,clip,width=.48\textwidth]
    {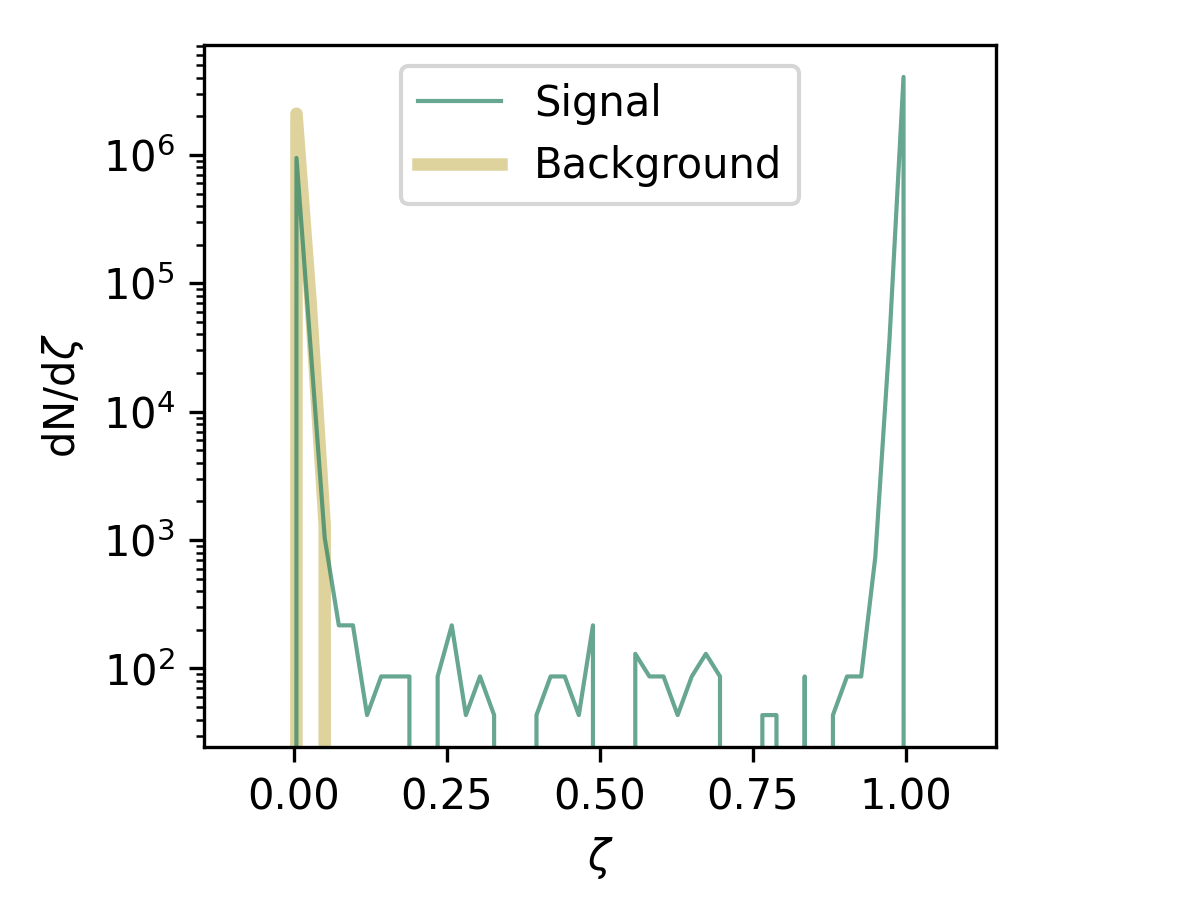}
    \includegraphics[trim=0mm 0mm 3mm 0mm,clip,width=.48\textwidth]{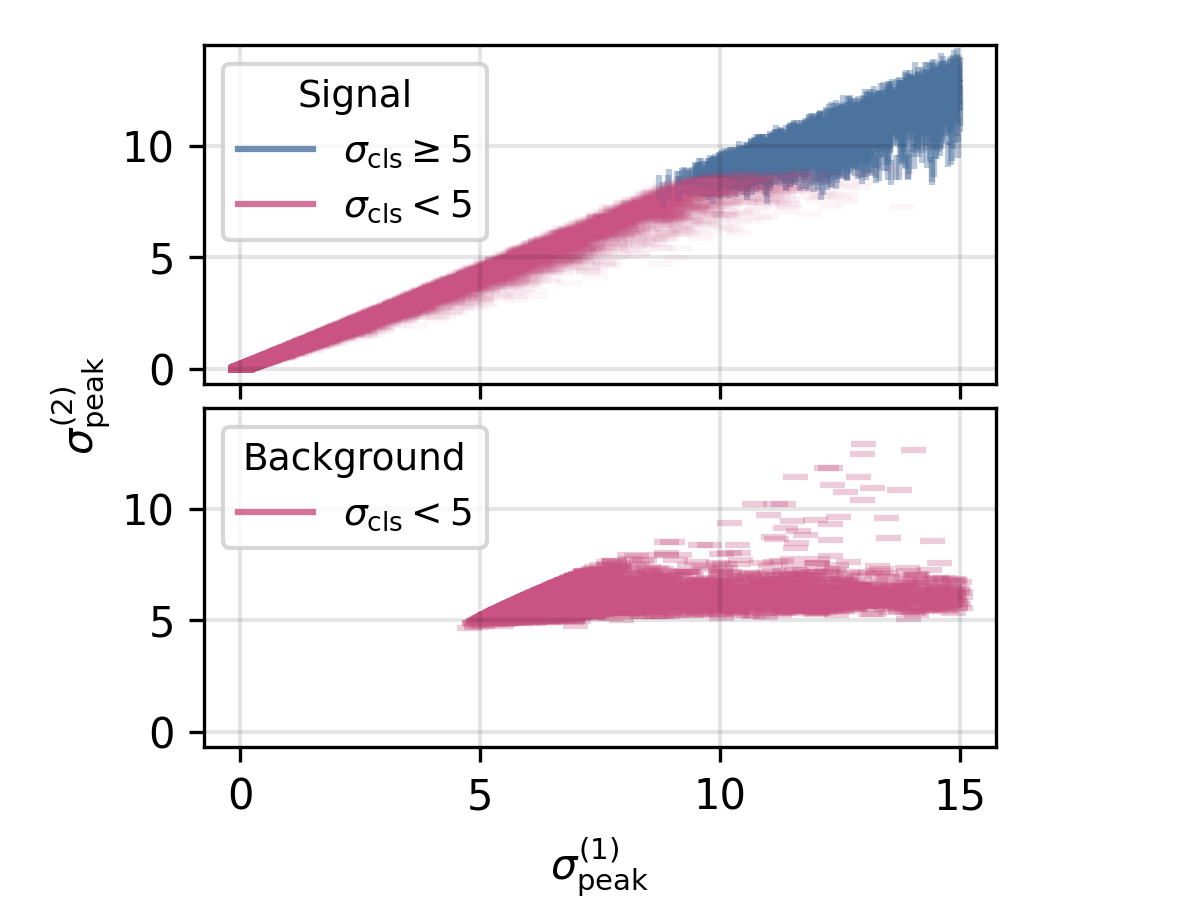}
    \caption{
        \textbf{Left:} Distributions 
        of the test
        statistic of the classification algorithm,
        $\zeta$, for the signal and background classes,
        as indicated. The background sample, corresponding
        to noise-based light curves, is clearly distinguished
        from most of the signal sample, which 
        corresponds to simulated flares of various brightness.
        The test statistic is mapped to a significance
        for source detection, denoted by $\sigma_{\mathrm{cls}}$.
        \textbf{Right:} Relation between the
        \snr of the brightest and 
        second-brightest 10~\secnd 
        interval of the light-curve,
        respectively denoted by 
        $\sigma_{\mathrm{peak}}^{(1)}$ and
        $\sigma_{\mathrm{peak}}^{(2)}$.
        For visual clarity, we only include 
        light curves for which 
        $\sigma_{\mathrm{peak}}^{(1,2)} \leq 15$
        per time step.
        The dataset is split into 
        categories for the signal and background
        samples, corresponding to light curves
        that pass or fail the $5\sigma$ detection
        threshold of the classifier.
        As indicated, background light curves
        and faint signal examples are not significantly
        detected. Above a certain threshold 
        (approximately $\sigma_{\mathrm{peak}}^{(1)} > 8$
        and $\sigma_{\mathrm{peak}}^{(2)} > 7$), 
        signal events
        may be identified with high confidence.
    \label{fig:clasif_ts_signifs}}
\end{figure*}
As indicated above, \grb flares would be detected
based on their light-curve. A simplistic approach
could be to identify events given two or more consecutive
data points that pass some signal-to-noise (\snr) threshold.
In practice, this strategy is suboptimal, and may
result in low detection efficiency or an overabundance 
of spurious detections. 

An alternative strategy is to devise
a test statistic, which encapsulates the signal-to-noise
of the entire light-curve, and can be mapped to
a \textit{p}-value (significance for source detection).
For this purpose,
we follow the approach developed by 
\cite{2020ApJ...894L..25S}. 
We construct
a toy model that illustrates the methodology.
Within the scope of the current
study we do not attempt to simulate 
the background to a realistic survey.
We also assume that stellar flares are identified
independently via a combination
of cross-matching with stellar catalogues;
a dedicated event classifier; and longer-term
follow-up.

We use 
the open-source software, 
\tensorflow~\citep{tensorflow2015-whitepaper},
to construct a simple
neural network. The network
is made up of four layers as follows:
\begin{enumerate*}[label=(\roman*)]
    \item two consecutive layers of
        $64$ and $32$~fully-connected
        neurons;
    \item a \textit{softmax} layer, which
        maps the outer dense layer to 
        a single number within the range, $[0, 1]$;
    \item a probabilistic layer, representing a
    normal distribution, which acts on the \textit{softmax}
    layer. The output of this layer, denoted by
    $\zeta$, serves as the test statistic
    for our analysis.
\end{enumerate*}

The inputs to the network are time series data,
constructed for a putative transient source position
over consecutive exposures.
For the wide observing-mode
segment of the light-curve, the
metric per time step is the
\snr for aperture photometry, which is derived
using the so-called ``CCD equation''
\citep{1989PASP..101..616H}.
We assume first detection and convergence of
the narrow subarray of telescopes 
after $30$~\secnd.
For this observation interval, the inputs
to the network correspond to
the integrated \snr from all telescopes.
Each light-curve
comprises $14$~steps, representing a total
of $150$~\secnd of observations.

The training objective for the network is to
perform classification between two classes
for ``background'' and ``signal''.
Background data correspond
to random noise, based on fluctuations
of the assumed sky brightness. 
We add to this  
further upward fluctuations on the level of
$(10 \pm 3)\sigma$, representing single-exposure
artefacts. These are injected randomly at a rate of about
one fluctuation per input time series.

Data comprising the signal class include simulations
of \grb optical flashes. We use the complete
unweighted sample, without accounting for the relative 
number density on $z$, \lumigamma, and \lumiratio.
Correspondingly, the training examples
are balanced with respect to flare
brightness. 
We apply a selection cut on
the signal class. The objective is
to suppress faint events that can not be distinguished
from background,
which can impede the training of the network. 
Explicitly, we impose the condition,
$\sigma_{\mathrm{peak}}^{(1)} > 12$ and
$\sigma_{\mathrm{peak}}^{(2)} > 8$. Here
$\sigma_{\mathrm{peak}}^{(1)}$ corresponds to
the \snr-input of the brightest 10~\secnd 
interval of the light-curve;
$\sigma_{\mathrm{peak}}^{(2)}$ likewise corresponds 
to the second-brightest interval.
This selection cut is only used for the 
training phase of the classification pipeline.

After the network is trained, the correspondence between
the output, $\zeta$, and the detection significance,
$\sigma_{\mathrm{cls}}$, is derived
numerically, as discussed by \cite{2020ApJ...894L..25S}. 
The performance of the classifier is shown
in Fig.~\ref{fig:clasif_ts_signifs}. 
We significantly detect flares and
reject background, based on the brightness
and the time structure of the signal.
For illustration, the detection threshold
is approximately characterised by flares having
$\sigma_{\mathrm{peak}}^{(1)} > 8$ and
$\sigma_{\mathrm{peak}}^{(2)} > 7$. 
However,
background realisations having brighter
fluctuations are also rejected. This is due
to the fact that the network
is trained to identify correlated
structures within the input time series.

\subsection{Detection rates} 
 
\begin{table*}[tp]
    \begin{center}
        \begin{threeparttable}
            \caption{\label{table:survey_summary}
                Summary of the parameters used to simulate
                the blind survey.
            }
        \begin{tabular}{cc}
            \textbf{GRB simulation parameters} &  \\
            \hline \hline
            Local event rate &
                $\rho{_0} = 1.3$~\gpcvolinv~\yrinv \\
            Redshift \& luminosity ranges & 
                $z \in [0, 6]$ \;,\; 
                $\log_{10}(\lumigamma\;[\lumiUnits]) \in [50, 54]$ \\
            \multirow{2}{*}{Temporal profile$^{\dag}$} 
            & 
                $\tau_{0} \in [5, 10]$ \;,\;
                $\tau_{1} \in [5, 20]$ \;,\; 
                $\tau_{2} \in [10, 50]~\secnd$; \\
            &
                $\alpha_{t} \in [0.2, 0.5]$ \;,\;
                $\beta_{t} \in [0.2, 0.3]$ \\
            Spectral index & 
                $\beta_{0} = 0.79 \pm 0.03$ \\
            Reference for \lumiratio calibration & 
                $\log_{10}(L_{\mathrm{R}}\;[\lumiUnits]) = 46.18 \pm 1.04$ \\
            Extinction & 
                $\log_{10}A_{\mathrm{V}} = -0.63 \pm 0.42$ \\
            \hline
            & \\
            \textbf{Survey parameters} &  \\
            \hline \hline
            Number of telescopes & 
                $20$~in wide mode, 
                or $20$~in narrow mode\\
            Intrinsic / effective survey area & 
                $150$ / $300$~\degsqr \\
            Single-exposure duration & 
                $2 \times \tau_{\mathrm{obs}} = 10$~\secnd \\
            Single-\fov cadence (wide mode) & 
                ${\text{1:2}}$ cadence with gaps, $\tau_{\mathrm{obs}} + 2 \times \tau_{\mathrm{slew}} = 20$~\secnd \\
            Effective live-time & 
                0.25~\yrinv (6~hours per night) \\
            \hline
            \end{tabular}
            \vspace{7pt}
            \begin{tablenotes}
                \item[$\dag$] 
                    Unlisted parameters, $a_{t}$; $b_{t}$; $c_{t}$; and
                    $d_{t}$, are derived on a case by case basis, given the luminosity normalisation of the light-curve.
            \end{tablenotes}
        \end{threeparttable}
    \end{center}
\end{table*}

The details of our simulation are summarised in
Table~\ref{table:survey_summary}.
For the moment, we assume that each and every 
\grb actually exhibits
an optical signal. In practice, this is not realistic,
as a substantial fraction of events should
be designated
as optically ``dark'' \grbs
\citep{2023Univ....9..113O}. 
We discuss this in Sec.~\ref{sec:summary_discussion}.

\begin{figure}[tp]
    \centering
    \includegraphics[trim=3mm 0mm 0mm 0mm,clip,width=.48\textwidth]{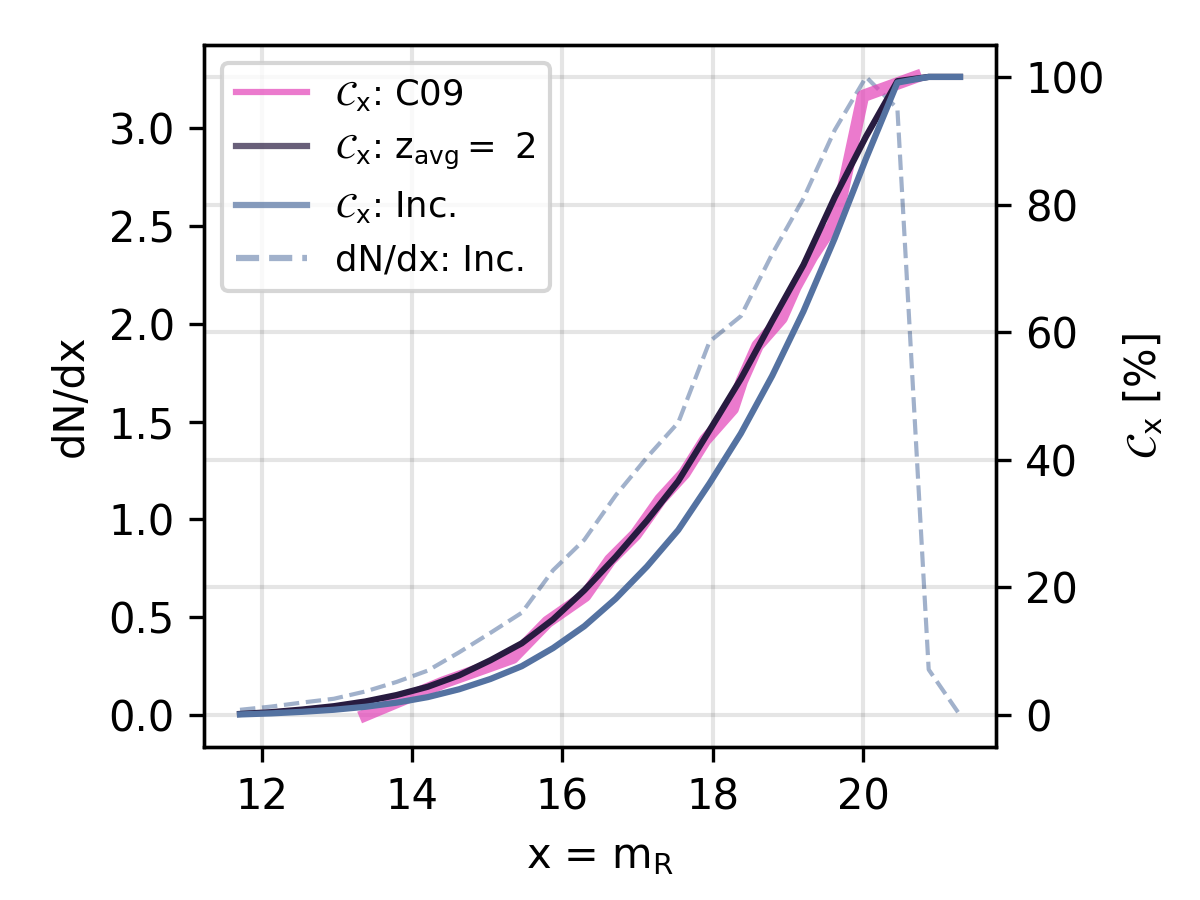}
    \caption{
        Comparison between the simulated \grb sample
        and the reference distribution from \cenkoA
        \citep{2009ApJ...693.1484C}.
        The dashed line represents the
        distribution of
        $R_{\mathrm{C}}$-band magnitudes,
        $\mathrm{m}_{\mathrm{R}}$, for our
        inclusive sample, indicated as ``Inc.''.
        The full lines represent the cumulative
        $\mathrm{m}_{\mathrm{R}}$ distributions from
        \cenkoA and for the current study,
        denoted by
        $\mathcal{C}_{\mathrm{x}}$. 
        Our results
        correspond either to a limited
        redshift range, for which the average is
        $z_{\mathrm{avg}} \sim 2$, or to the inclusive
        dataset, as indicated. 
        The limited redshift sample
        is comparable to the equivalent reference
        data, validating the luminosity
        calibration procedure.
    \label{fig:cumulative_dist_t_1000}}
\end{figure}

We begin by verifying 
our predictions for the reference luminosity 
distribution of \cenkoA,
as shown in Fig.~\ref{fig:cumulative_dist_t_1000}.
The expected distribution of observed magnitudes
is recovered
for a comparable volumetric sample
(average redshift, $z_{\mathrm{avg}} \sim 2$).
For our full dataset the 
distribution is fainter, as it extends to higher redshift.
We find the following cumulative fractions,
${\mathcal{C} < \mathrm{m}_{\mathrm{inc}}}$,
at $\powB{3}$~\secnd:
$5\% < 14.5$, $20\% < 17$, and $50\% < 18.5$.

\begin{figure*}[tp]
    \centering
    \includegraphics[trim=3mm 0mm 0mm 0mm,clip,width=.48\textwidth]
    {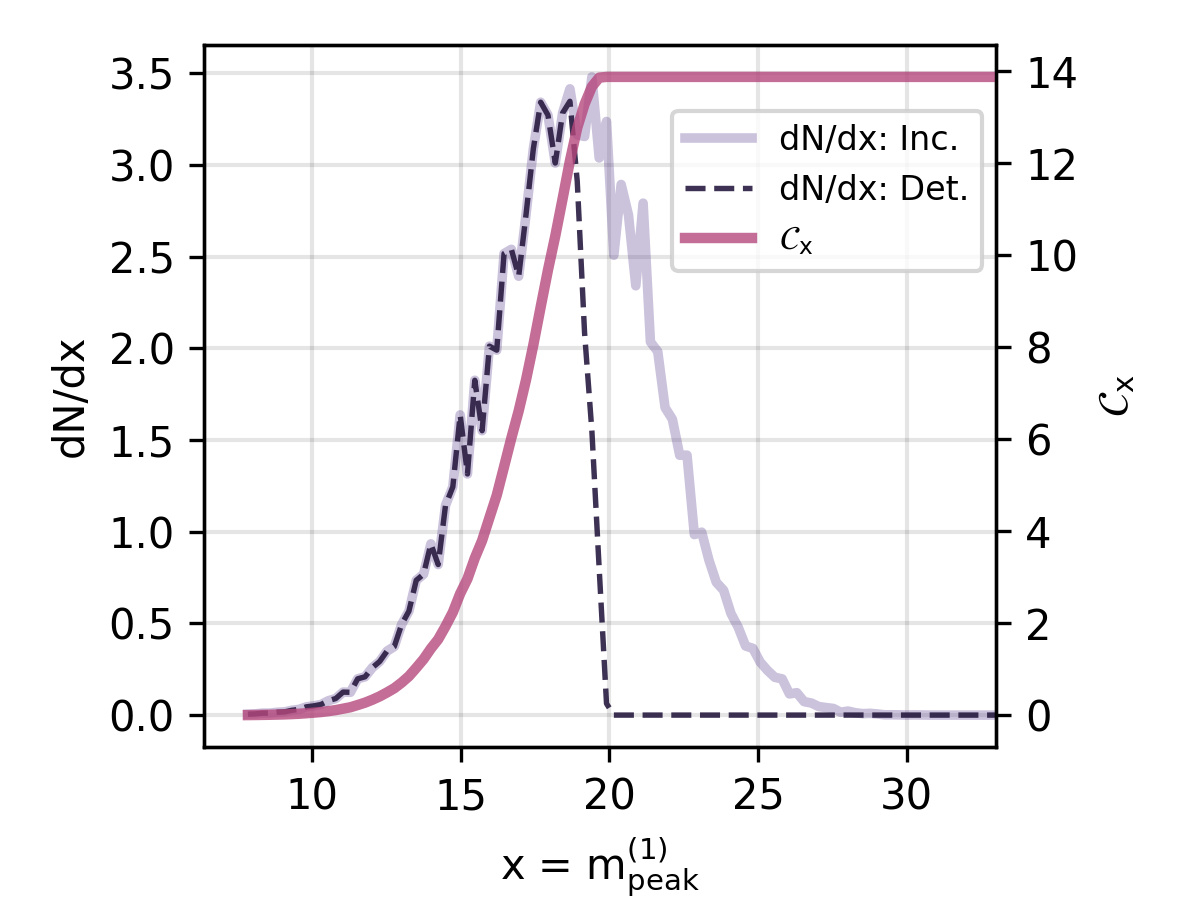}
    \includegraphics[trim=3mm 0mm 0mm 0mm,clip,width=.48\textwidth]{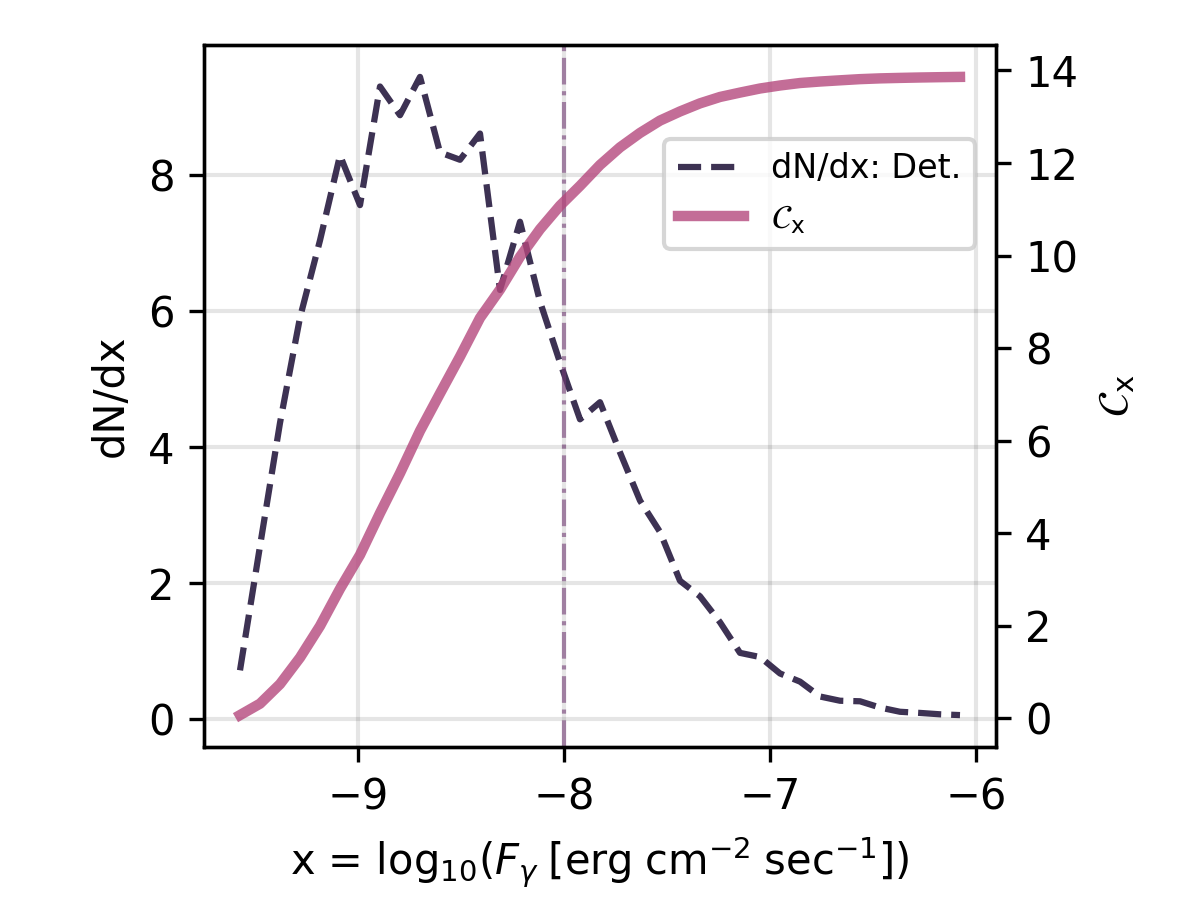}
        \caption{
        Simulated \grbs, described by two
        metrics, denoted by $\mathrm{x}$.
        In total, about $24$~\grbs~\yrinv would
        occur within the \fov of the survey,
        $14$ of which are detectable.
        Out of these,
        three events are also expected to
        trigger \he instruments. (This result
        does not account for a suppression of
        the optical flux for dark \grbs, which
        reduces the expected detectable rate
        from~14 to 7--11~\grbs~\yrinv, as discussed
        in the text.)
        \textbf{Left:} Distributions, $\mathrm{dN/dx}$, of
        the peak light-curve magnitude,
        $\mathrm{m}_{\mathrm{peak}}^{(1)}$, of \last.
        The different curves represent
        the full simulated sample of events
        ("Inc."), and
        the significantly detected subsample
        ("Det.").
        Also shown is the respective 
        cumulative distribution of detected
        events, $\mathcal{C}_{x}$,
        as indicated.
        \textbf{Right:} Distribution, $\mathrm{dN/dx}$, 
        and cumulative distribution, $\mathcal{C}_{x}$, 
        of the
        peak \gamray flux, $F_{\gamma}$,
        for detected \grbs, as indicated.
        The
        typical threshold of \gamray satellites, 
        $F_{\gamma} \sim \powB{-8}$~\fluxUnits,
        is highlighted by the dotted-dashed vertical line.
    \label{fig:detected_mag_dist}}
\end{figure*}

Distributions of detectable flares
are presented in Fig.~\ref{fig:detected_mag_dist}.
We impose a minimal selection threshold, where
light curves
include at least two time steps (10~\secnd exposures)
having $\snr > 5$. In addition, we require
a classification detection significance,
$\sigma_{\mathrm{cls}} > 5$.
For the assumed wide subarray of
$20$~telescopes, operating with a ${\text{1:2}}$~cadence,
the effective survey area is approximately $300$~\degsqr.
We further assume the availability of
6~hours of clear sky per night,
corresponding to an effective survey live-time
of $0.25$~\yrinv
\citep{2023PASP..135f5001O}.

Events are observable up to peak magnitudes,
$\mathrm{m}_{\mathrm{peak}}^{(1)} \sim 19.5$,
over 10~\secnd exposures.
In total, this survey strategy allows 
detection of up to about $14$~\grbs~\yrinv during
their early emission phase.
Assuming an \he flux threshold,
$F_{\gamma} > \powB{-8}$~\fluxUnits,
a subset of 3~\grbs~\yrinv is
detectable. These events
would also trigger \gamray satellites 
observing the same field as \last.
The rest of the events are not expected
to be associated with \he triggers, but might
be detectable via follow-up archival search.
In reality, the expected rate
of detectable \grbs is actually lower
than~14~\grbs~\yrinv,
as discussed in next section.

\section{Summary and discussion} 
\label{sec:summary_discussion}

We suggest
an observational approach
for the independent discovery of long
\grbs. Events are detected via
their early
optical signals, with the purpose
of probing the prompt phase of the
emission when relevant. 
We explore
the potential of an
array of small telescopes such as \last for
this purpose. 
Such an instrument can be used
to scan a wide region of the sky, and then
focus on a smaller FoV with increased 
sensitivity.

We find that for our chosen
array configuration,
one can potentially detect
$14$~\grbs~\yrinv. Of these, 
$20\%$ would also
be detectable with high confidence by
\gamray instruments.
An important caveat for this result is 
our assumptions on the realistic fraction of
optically detectable \grbs, given
that some events lack optical emission.
A \grb is commonly classified as optically ``dark'' when 
the ratio of optical to X-ray flux is incompatible
with the standard expectations
for synchrotron afterglows\footnote{The optical $R_{\mathrm{C}}$
to X-ray spectral index of dark \grbs under this definition satisfies 
$\beta_{\mathrm{opt/X}} < 0.5$.} \citep{2004ApJ...617L..21J}.

The fraction of dark \grbs is not straightforward to estimate.
Several factors may play a role
in masking the observed signal. These include extrinsic
effects, such as Lyman~$\alpha$ absorption at high redshift,
or high host galaxy extinction.
Intrinsic suppression within the source and its environment
may also play an important role. The time at which
the flux in the optical and X-ray bands is compared is
also determinative, and can \eg be biased by 
continuous or renewed activity of the central engine.

We note that, under this definition, the optical signal
of a dark \grb may yet be observed; however the flux
is significantly suppressed compared to other bands.
To put this in context, \cenkoA 
optically detected $\sim80\%$
of their sample
of 29~\grbs, for which follow-up began within
an hour of the trigger. The fraction
of bursts classified as dark for the same dataset
was found to be $\sim50\%$.
The detection fraction inferred from
Fig.~\ref{fig:detected_mag_dist}
may therefore be more realistically estimated as
7--11~\grbs~\yrinv. It is reasonable to assume
that the same number of
events, 3~\grbs~\yrinv, would have corresponding \he triggers.
These would be relatively bright events, 
for which stronger optical components are
expected
\citep{2017Natur.547..425T}. 
However, the details depend on 
whether the nature of
a dark burst is intrinsic or extrinsic.
These numbers also depend on the assumption that
it is valid to model the optical flash
as synchrotron emission, which may not be the
case for every \grb.

The focus of this work is the
prompt optical emission of \grbs, for which few
data exist as reference.
Our simulations are therefore based on
early time afterglow data, where we extrapolate the 
emission from \powB{3}~\secnd
back
to the moment of the explosion.
The temporal decay of \grb afterglow flux at late times
(hours to days in this context) 
is very generally compatible 
with~$f \propto t^{-1}$
\citep{2010ApJ...720.1513K}.
We purposely choose shallower profiles at 
earlier times for
the current study, so that we do
not overestimate the flux; this
is also broadly compatible with
available observations
(\eg \cite{2012ApJ...758...27L, 2013ApJ...772...73K}).
We note that it is possible 
that an additional component
is present in some events, directly attributed to
the activity of the engine. Detecting this
component is in fact the main motivation
for future surveys.
In this context, our predictions
for the brightness of the early time signal
may be considered conservative.

Previous studies involving speedy \grb follow-up
are mostly of limited sample-size,
and are generally
biased towards early-afterglows. 
While such observations are 
not necessarily our desired point of reference, 
it remains interesting to compare them
with our predictions.
We find the following cumulative fractions,
${\mathcal{C} < \mathrm{m}_{\mathrm{inc}}}$,
during the prompt phase:
$2\% < 12$, $9\% < 14$,  $30\% < 16$,
$50\% < 17.2$, and $90\% < 19$.
These are comparable with
previous findings,
\eg those of \cite{2009AJ....137.4100K, 2013ApJ...774..132W}.

We also verify that our results are compatible with
established limits on the
rate of fast extragalactic transients.
For example, ROTSE-III, found
$\mathcal{R}_{\mathrm{\grb}} < 1.9$~\degsqrinv~\yrinv
up to $m_{\mathrm{R}} \sim 18$
\citep{2005ApJ...631.1032R},
and MASTER derived a comparable limit,
$\mathcal{R}_{\mathrm{\grb}} < 1.2$~\degsqrinv~\yrinv up 
to $m_{\mathrm{V}} = 17.5$ by 
\citep{2007ARep...51.1004L}.
(See also Table~3 of \cite{2013ApJ...779...18B};
Fig.~6 of
\cite{2020MNRAS.491.5852A}.)
Accounting for survey coverage, live-time, and
a realistic fraction of optically dark
events,
the current study finds an intrinsic rate of
$\mathcal{R}_{\mathrm{\grb}} \sim 0.12$~\degsqrinv~\yrinv.
This result is consistent
with the previous nondetections; it
illustrates
that an order of magnitude improvement
in efficiency is required in order
to make progress in the field.
This highlights the importance of conducting
very wide surveys, coupled to analyses
targeting short time scale transients.

\section{Acknowledgments} 

We would like to thank the following individuals for
numerous fruitful discussions in support
of this work:
S.~Ben-Ami,
R.~Buehler,
J.~Borowska,
V.~Fallah Ramazani,
A.~Franckowiak,
S.~Garrappa,
O.~Gueta,
A.~Y.~Q.~Ho,
R.~Konno,
E.~O.~Ofek,
P.~Rekhi,
J.~Sinapius
N.~L.~Strotjohann, and
S.~Weimann.
We would also like to thank the anonymous 
referee for their insightful comments.

\bibliography{main}{}

\begin{thebibliography}{}
\expandafter\ifx\csname natexlab\endcsname\relax\def\natexlab#1{#1}\fi
\providecommand{\url}[1]{\href{#1}{#1}}
\providecommand{\dodoi}[1]{doi:~\href{http://doi.org/#1}{\nolinkurl{#1}}}
\providecommand{\doeprint}[1]{\href{http://ascl.net/#1}{\nolinkurl{http://ascl.net/#1}}}
\providecommand{\doarXiv}[1]{\href{https://arxiv.org/abs/#1}{\nolinkurl{https://arxiv.org/abs/#1}}}

\bibitem[{Abadi {et~al.}(2015)Abadi, Agarwal, Barham, Brevdo, Chen, Citro,
  Corrado, Davis, Dean, Devin, Ghemawat, Goodfellow, Harp, Irving, Isard, Jia,
  Jozefowicz, Kaiser, Kudlur, Levenberg, Man\'{e}, Monga, Moore, Murray, Olah,
  Schuster, Shlens, Steiner, Sutskever, Talwar, Tucker, Vanhoucke, Vasudevan,
  Vi\'{e}gas, Vinyals, Warden, Wattenberg, Wicke, Yu, \&
  Zheng}]{tensorflow2015-whitepaper}
Abadi, M., Agarwal, A., Barham, P., {et~al.} 2015.
\newblock \url{https://www.tensorflow.org/}

\bibitem[{{Aizawa} {et~al.}(2022){Aizawa}, {Kawana}, {Kashiyama}, {Ohsawa},
  {Kawahara}, {Naokawa}, {Tajiri}, {Arima}, {Jiang}, {Hartwig}, {Fujisawa},
  {Shigeyama}, {Arimatsu}, {Doi}, {Kasuga}, {Kobayashi}, {Kondo}, {Mori},
  {Okumura}, {Takita}, \& {Sako}}]{2022PASJ...74.1069A}
{Aizawa}, M., {Kawana}, K., {Kashiyama}, K., {et~al.} 2022, \pasj, 74, 1069,
  \dodoi{10.1093/pasj/psac056}

\bibitem[{{Andreoni} {et~al.}(2020{\natexlab{a}}){Andreoni}, {Cooke}, {Webb},
  {Rest}, {Pritchard}, {Caleb}, {Chang}, {Farah}, {Lien}, {M{\"o}ller},
  {Ravasio}, {Abbott}, {Bhandari}, {Cucchiara}, {Flynn}, {Jankowski}, {Keane},
  {Moriya}, {Onken}, {Parthasarathy}, {Price}, {Petroff}, {Ryder}, {Vohl}, \&
  {Wolf}}]{2020MNRAS.491.5852A}
{Andreoni}, I., {Cooke}, J., {Webb}, S., {et~al.} 2020{\natexlab{a}}, \mnras,
  491, 5852, \dodoi{10.1093/mnras/stz3381}

\bibitem[{{Andreoni} {et~al.}(2020{\natexlab{b}}){Andreoni}, {Lu}, {Smith},
  {Masci}, {Bellm}, {Graham}, {Kaplan}, {Kasliwal}, {Kaye}, {Kupfer}, {Laher},
  {Mahabal}, {Nordin}, {Porter}, {Prince}, {Reiley}, {Riddle}, {Van Roestel},
  \& {Yao}}]{2020ApJ...896L...2A}
{Andreoni}, I., {Lu}, W., {Smith}, R.~M., {et~al.} 2020{\natexlab{b}}, \apjl,
  896, L2, \dodoi{10.3847/2041-8213/ab94a5}

\bibitem[{{Andreoni} {et~al.}(2021){Andreoni}, {Coughlin}, {Kool}, {Kasliwal},
  {Kumar}, {Bhalerao}, {Carracedo}, {Ho}, {Pang}, {Saraogi}, {Sharma},
  {Shenoy}, {Burns}, {Ahumada}, {Anand}, {Singer}, {Perley}, {De}, {Fremling},
  {Bellm}, {Bulla}, {Crellin-Quick}, {Dietrich}, {Drake}, {Duev}, {Goobar},
  {Graham}, {Kaplan}, {Kulkarni}, {Laher}, {Mahabal}, {Shupe}, {Sollerman},
  {Walters}, \& {Yao}}]{2021ApJ...918...63A}
{Andreoni}, I., {Coughlin}, M.~W., {Kool}, E.~C., {et~al.} 2021, \apj, 918, 63,
  \dodoi{10.3847/1538-4357/ac0bc7}

\bibitem[{{Arimatsu} {et~al.}(2021){Arimatsu}, {Tsumura}, {Usui}, {Ootsubo}, \&
  {Watanabe}}]{2021AJ....161..135A}
{Arimatsu}, K., {Tsumura}, K., {Usui}, F., {Ootsubo}, T., \& {Watanabe}, J.-i.
  2021, \aj, 161, 135, \dodoi{10.3847/1538-3881/abd94d}

\bibitem[{{Becerra} {et~al.}(2021){Becerra}, {De Colle}, {Cant{\'o}}, {Lizano},
  {Gonz{\'a}lez}, {Granot}, {Klotz}, {Watson}, {Fraija}, {Araudo}, {Troja},
  {Atteia}, {Lee}, {Turpin}, {Bloom}, {Boer}, {Butler}, {Gonz{\'a}lez},
  {Kutyrev}, {Prochaska}, {Ramirez-Ruiz}, {Richer}, \&
  {Rom{\'a}n-Z{\'u}{\~n}iga}}]{2021ApJ...908...39B}
{Becerra}, R.~L., {De Colle}, F., {Cant{\'o}}, J., {et~al.} 2021, \apj, 908,
  39, \dodoi{10.3847/1538-4357/abcd3a}

\bibitem[{{Ben-Ami} {et~al.}(2023){Ben-Ami}, {Ofek}, {Polishook},
  {Franckowiak}, {Hallakoun}, {Segre}, {Shvartzvald}, {Strotjohann}, {Yaron},
  {Aharonson}, {Arcavi}, {Berge}, {Ramazani}, {Gal-Yam}, {Garrappa}, {Hershko},
  {Nir}, {Ohm}, {Rybicki}, {Sadeh}, {Segev}, {Shani}, {Sofer-Rimalt}, \&
  {Weimann}}]{2023PASP..135h5002B}
{Ben-Ami}, S., {Ofek}, E.~O., {Polishook}, D., {et~al.} 2023, \pasp, 135,
  085002, \dodoi{10.1088/1538-3873/aceb30}

\bibitem[{{Berger} {et~al.}(2013){Berger}, {Leibler}, {Chornock}, {Rest},
  {Foley}, {Soderberg}, {Price}, {Burgett}, {Chambers}, {Flewelling}, {Huber},
  {Magnier}, {Metcalfe}, {Stubbs}, \& {Tonry}}]{2013ApJ...779...18B}
{Berger}, E., {Leibler}, C.~N., {Chornock}, R., {et~al.} 2013, \apj, 779, 18,
  \dodoi{10.1088/0004-637X/779/1/18}

\bibitem[{{Boubert} \& {Everall}(2020)}]{2020MNRAS.497.4246B}
{Boubert}, D., \& {Everall}, A. 2020, \mnras, 497, 4246,
  \dodoi{10.1093/mnras/staa2305}

\bibitem[{{Bromberg} {et~al.}(2011){Bromberg}, {Nakar}, \&
  {Piran}}]{2011ApJ...739L..55B}
{Bromberg}, O., {Nakar}, E., \& {Piran}, T. 2011, \apjl, 739, L55,
  \dodoi{10.1088/2041-8205/739/2/L55}

\bibitem[{{Cano} {et~al.}(2017){Cano}, {Wang}, {Dai}, \&
  {Wu}}]{2017AdAst2017E...5C}
{Cano}, Z., {Wang}, S.-Q., {Dai}, Z.-G., \& {Wu}, X.-F. 2017, Advances in
  Astronomy, 2017, 8929054, \dodoi{10.1155/2017/8929054}

\bibitem[{{Cenko} {et~al.}(2009){Cenko}, {Kelemen}, {Harrison}, {Fox},
  {Kulkarni}, {Kasliwal}, {Ofek}, {Rau}, {Gal-Yam}, {Frail}, \&
  {Moon}}]{2009ApJ...693.1484C}
{Cenko}, S.~B., {Kelemen}, J., {Harrison}, F.~A., {et~al.} 2009, \apj, 693,
  1484, \dodoi{10.1088/0004-637X/693/2/1484}

\bibitem[{{Cenko} {et~al.}(2013){Cenko}, {Kulkarni}, {Horesh}, {Corsi}, {Fox},
  {Carpenter}, {Frail}, {Nugent}, {Perley}, {Gruber}, {Gal-Yam}, {Groot},
  {Hallinan}, {Ofek}, {Rau}, {MacLeod}, {Miller}, {Bloom}, {Filippenko},
  {Kasliwal}, {Law}, {Morgan}, {Polishook}, {Poznanski}, {Quimby}, {Sesar},
  {Shen}, {Silverman}, \& {Sternberg}}]{2013ApJ...769..130C}
{Cenko}, S.~B., {Kulkarni}, S.~R., {Horesh}, A., {et~al.} 2013, \apj, 769, 130,
  \dodoi{10.1088/0004-637X/769/2/130}

\bibitem[{{Cenko} {et~al.}(2015){Cenko}, {Urban}, {Perley}, {Horesh}, {Corsi},
  {Fox}, {Cao}, {Kasliwal}, {Lien}, {Arcavi}, {Bloom}, {Butler}, {Cucchiara},
  {de Diego}, {Filippenko}, {Gal-Yam}, {Gehrels}, {Georgiev}, {Jes{\'u}s
  Gonz{\'a}lez}, {Graham}, {Greiner}, {Kann}, {Klein}, {Knust}, {Kulkarni},
  {Kutyrev}, {Laher}, {Lee}, {Nugent}, {Prochaska}, {Ramirez-Ruiz}, {Richer},
  {Rubin}, {Urata}, {Varela}, {Watson}, \& {Wozniak}}]{2015ApJ...803L..24C}
{Cenko}, S.~B., {Urban}, A.~L., {Perley}, D.~A., {et~al.} 2015, \apjl, 803,
  L24, \dodoi{10.1088/2041-8205/803/2/L24}

\bibitem[{{Corsi} {et~al.}(2023){Corsi}, {Ho}, {Cenko}, {Kulkarni}, {Anand},
  {Yang}, {Sollerman}, {Srinivasaragavan}, {Omand}, {Balasubramanian}, {Frail},
  {Fremling}, {Perley}, {Yao}, {Dahiwale}, {De}, {Dugas}, {Hankins}, {Jencson},
  {Kasliwal}, {Tzanidakis}, {Bellm}, {Laher}, {Masci}, {Purdum}, \&
  {Regnault}}]{2023ApJ...953..179C}
{Corsi}, A., {Ho}, A. Y.~Q., {Cenko}, S.~B., {et~al.} 2023, \apj, 953, 179,
  \dodoi{10.3847/1538-4357/acd3f2}

\bibitem[{{Dainotti} {et~al.}(2020){Dainotti}, {Livermore}, {Kann}, {Li},
  {Oates}, {Yi}, {Zhang}, {Gendre}, {Cenko}, \& {Fraija}}]{2020ApJ...905L..26D}
{Dainotti}, M.~G., {Livermore}, S., {Kann}, D.~A., {et~al.} 2020, \apjl, 905,
  L26, \dodoi{10.3847/2041-8213/abcda9}

\bibitem[{{Dyer} {et~al.}(2018){Dyer}, {Dhillon}, {Littlefair}, {Steeghs},
  {Ulaczyk}, {Chote}, {Galloway}, \& {Rol}}]{2018SPIE10704E..0CD}
{Dyer}, M.~J., {Dhillon}, V.~S., {Littlefair}, S., {et~al.} 2018, in Society of
  Photo-Optical Instrumentation Engineers (SPIE) Conference Series, Vol. 10704,
  Observatory Operations: Strategies, Processes, and Systems VII, 107040C,
  \dodoi{10.1117/12.2311865}

\bibitem[{{Eichler} {et~al.}(1989){Eichler}, {Livio}, {Piran}, \&
  {Schramm}}]{1989Natur.340..126E}
{Eichler}, D., {Livio}, M., {Piran}, T., \& {Schramm}, D.~N. 1989, \nat, 340,
  126, \dodoi{10.1038/340126a0}

\bibitem[{{Gaia Collaboration} {et~al.}(2016){Gaia Collaboration}, {Prusti},
  {de Bruijne}, {Brown}, {Vallenari}, {Babusiaux}, {Bailer-Jones}, {Bastian},
  {Biermann}, {Evans}, {Eyer}, {Jansen}, {Jordi}, {Klioner}, {Lammers},
  {Lindegren}, {Luri}, {Mignard}, {Milligan}, {Panem}, {Poinsignon},
  {Pourbaix}, {Randich}, {Sarri}, {Sartoretti}, {Siddiqui}, {Soubiran},
  {Valette}, {van Leeuwen}, {Walton}, {Aerts}, {Arenou}, {Cropper}, {Drimmel},
  {H{\o}g}, {Katz}, {Lattanzi}, {O'Mullane}, {Grebel}, {Holland}, {Huc},
  {Passot}, {Bramante}, {Cacciari}, {Casta{\~n}eda}, {Chaoul}, {Cheek}, {De
  Angeli}, {Fabricius}, {Guerra}, {Hern{\'a}ndez}, {Jean-Antoine-Piccolo},
  {Masana}, {Messineo}, {Mowlavi}, {Nienartowicz}, {Ord{\'o}{\~n}ez-Blanco},
  {Panuzzo}, {Portell}, {Richards}, {Riello}, {Seabroke}, {Tanga},
  {Th{\'e}venin}, {Torra}, {Els}, {Gracia-Abril}, {Comoretto},
  {Garcia-Reinaldos}, {Lock}, {Mercier}, {Altmann}, {Andrae}, {Astraatmadja},
  {Bellas-Velidis}, {Benson}, {Berthier}, {Blomme}, {Busso}, {Carry},
  {Cellino}, {Clementini}, {Cowell}, {Creevey}, {Cuypers}, {Davidson}, {De
  Ridder}, {de Torres}, {Delchambre}, {Dell'Oro}, {Ducourant}, {Fr{\'e}mat},
  {Garc{\'\i}a-Torres}, {Gosset}, {Halbwachs}, {Hambly}, {Harrison}, {Hauser},
  {Hestroffer}, {Hodgkin}, {Huckle}, {Hutton}, {Jasniewicz}, {Jordan},
  {Kontizas}, {Korn}, {Lanzafame}, {Manteiga}, {Moitinho}, {Muinonen},
  {Osinde}, {Pancino}, {Pauwels}, {Petit}, {Recio-Blanco}, {Robin}, {Sarro},
  {Siopis}, {Smith}, {Smith}, {Sozzetti}, {Thuillot}, {van Reeven}, {Viala},
  {Abbas}, {Abreu Aramburu}, {Accart}, {Aguado}, {Allan}, {Allasia},
  {Altavilla}, {{\'A}lvarez}, {Alves}, {Anderson}, {Andrei}, {Anglada Varela},
  {Antiche}, {Antoja}, {Ant{\'o}n}, {Arcay}, {Atzei}, {Ayache}, {Bach},
  {Baker}, {Balaguer-N{\'u}{\~n}ez}, {Barache}, {Barata}, {Barbier}, {Barblan},
  {Baroni}, {Barrado y Navascu{\'e}s}, {Barros}, {Barstow}, {Becciani},
  {Bellazzini}, {Bellei}, {Bello Garc{\'\i}a}, {Belokurov}, {Bendjoya},
  {Berihuete}, {Bianchi}, {Bienaym{\'e}}, {Billebaud}, {Blagorodnova},
  {Blanco-Cuaresma}, {Boch}, {Bombrun}, {Borrachero}, {Bouquillon}, {Bourda},
  {Bouy}, {Bragaglia}, {Breddels}, {Brouillet}, {Br{\"u}semeister},
  {Bucciarelli}, {Budnik}, {Burgess}, {Burgon}, {Burlacu}, {Busonero}, {Buzzi},
  {Caffau}, {Cambras}, {Campbell}, {Cancelliere}, {Cantat-Gaudin}, {Carlucci},
  {Carrasco}, {Castellani}, {Charlot}, {Charnas}, {Charvet}, {Chassat},
  {Chiavassa}, {Clotet}, {Cocozza}, {Collins}, {Collins}, {Costigan}, {Crifo},
  {Cross}, {Crosta}, {Crowley}, {Dafonte}, {Damerdji}, {Dapergolas}, {David},
  {David}, {De Cat}, {de Felice}, {de Laverny}, {De Luise}, {De March}, {de
  Martino}, {de Souza}, {Debosscher}, {del Pozo}, {Delbo}, {Delgado},
  {Delgado}, {di Marco}, {Di Matteo}, {Diakite}, {Distefano}, {Dolding}, {Dos
  Anjos}, {Drazinos}, {Dur{\'a}n}, {Dzigan}, {Ecale}, {Edvardsson}, {Enke},
  {Erdmann}, {Escolar}, {Espina}, {Evans}, {Eynard Bontemps}, {Fabre},
  {Fabrizio}, {Faigler}, {Falc{\~a}o}, {Farr{\`a}s Casas}, {Faye}, {Federici},
  {Fedorets}, {Fern{\'a}ndez-Hern{\'a}ndez}, {Fernique}, {Fienga}, {Figueras},
  {Filippi}, {Findeisen}, {Fonti}, {Fouesneau}, {Fraile}, {Fraser}, {Fuchs},
  {Furnell}, {Gai}, {Galleti}, {Galluccio}, {Garabato}, {Garc{\'\i}a-Sedano},
  {Gar{\'e}}, {Garofalo}, {Garralda}, {Gavras}, {Gerssen}, {Geyer}, {Gilmore},
  {Girona}, {Giuffrida}, {Gomes}, {Gonz{\'a}lez-Marcos},
  {Gonz{\'a}lez-N{\'u}{\~n}ez}, {Gonz{\'a}lez-Vidal}, {Granvik}, {Guerrier},
  {Guillout}, {Guiraud}, {G{\'u}rpide}, {Guti{\'e}rrez-S{\'a}nchez}, {Guy},
  {Haigron}, {Hatzidimitriou}, {Haywood}, {Heiter}, {Helmi}, {Hobbs},
  {Hofmann}, {Holl}, {Holland}, {Hunt}, {Hypki}, {Icardi}, {Irwin}, {Jevardat
  de Fombelle}, {Jofr{\'e}}, {Jonker}, {Jorissen}, {Julbe}, {Karampelas},
  {Kochoska}, {Kohley}, {Kolenberg}, {Kontizas}, {Koposov}, {Kordopatis},
  {Koubsky}, {Kowalczyk}, {Krone-Martins}, {Kudryashova}, {Kull}, {Bachchan},
  {Lacoste-Seris}, {Lanza}, {Lavigne}, {Le Poncin-Lafitte}, {Lebreton},
  {Lebzelter}, {Leccia}, {Leclerc}, {Lecoeur-Taibi}, {Lemaitre}, {Lenhardt},
  {Leroux}, {Liao}, {Licata}, {Lindstr{\o}m}, {Lister}, {Livanou}, {Lobel},
  {L{\"o}ffler}, {L{\'o}pez}, {Lopez-Lozano}, {Lorenz}, {Loureiro},
  {MacDonald}, {Magalh{\~a}es Fernandes}, {Managau}, {Mann}, {Mantelet},
  {Marchal}, {Marchant}, {Marconi}, {Marie}, {Marinoni}, {Marrese},
  {Marschalk{\'o}}, {Marshall}, {Mart{\'\i}n-Fleitas}, {Martino}, {Mary},
  {Matijevi{\v{c}}}, {Mazeh}, {McMillan}, {Messina}, {Mestre}, {Michalik},
  {Millar}, {Miranda}, {Molina}, {Molinaro}, {Molinaro}, {Moln{\'a}r},
  {Moniez}, {Montegriffo}, {Monteiro}, {Mor}, {Mora}, {Morbidelli}, {Morel},
  {Morgenthaler}, {Morley}, {Morris}, {Mulone}, {Muraveva}, {Musella},
  {Narbonne}, {Nelemans}, {Nicastro}, {Noval}, {Ord{\'e}novic},
  {Ordieres-Mer{\'e}}, {Osborne}, {Pagani}, {Pagano}, {Pailler}, {Palacin},
  {Palaversa}, {Parsons}, {Paulsen}, {Pecoraro}, {Pedrosa}, {Pentik{\"a}inen},
  {Pereira}, {Pichon}, {Piersimoni}, {Pineau}, {Plachy}, {Plum}, {Poujoulet},
  {Pr{\v{s}}a}, {Pulone}, {Ragaini}, {Rago}, {Rambaux}, {Ramos-Lerate},
  {Ranalli}, {Rauw}, {Read}, {Regibo}, {Renk}, {Reyl{\'e}}, {Ribeiro},
  {Rimoldini}, {Ripepi}, {Riva}, {Rixon}, {Roelens}, {Romero-G{\'o}mez},
  {Rowell}, {Royer}, {Rudolph}, {Ruiz-Dern}, {Sadowski}, {Sagrist{\`a}
  Sell{\'e}s}, {Sahlmann}, {Salgado}, {Salguero}, {Sarasso}, {Savietto},
  {Schnorhk}, {Schultheis}, {Sciacca}, {Segol}, {Segovia}, {Segransan},
  {Serpell}, {Shih}, {Smareglia}, {Smart}, {Smith}, {Solano}, {Solitro},
  {Sordo}, {Soria Nieto}, {Souchay}, {Spagna}, {Spoto}, {Stampa}, {Steele},
  {Steidelm{\"u}ller}, {Stephenson}, {Stoev}, {Suess}, {S{\"u}veges}, {Surdej},
  {Szabados}, {Szegedi-Elek}, {Tapiador}, {Taris}, {Tauran}, {Taylor},
  {Teixeira}, {Terrett}, {Tingley}, {Trager}, {Turon}, {Ulla}, {Utrilla},
  {Valentini}, {van Elteren}, {Van Hemelryck}, {van Leeuwen}, {Varadi},
  {Vecchiato}, {Veljanoski}, {Via}, {Vicente}, {Vogt}, {Voss}, {Votruba},
  {Voutsinas}, {Walmsley}, {Weiler}, {Weingrill}, {Werner}, {Wevers},
  {Whitehead}, {Wyrzykowski}, {Yoldas}, {{\v{Z}}erjal}, {Zucker}, {Zurbach},
  {Zwitter}, {Alecu}, {Allen}, {Allende Prieto}, {Amorim},
  {Anglada-Escud{\'e}}, {Arsenijevic}, {Azaz}, {Balm}, {Beck}, {Bernstein},
  {Bigot}, {Bijaoui}, {Blasco}, {Bonfigli}, {Bono}, {Boudreault}, {Bressan},
  {Brown}, {Brunet}, {Bunclark}, {Buonanno}, {Butkevich}, {Carret}, {Carrion},
  {Chemin}, {Ch{\'e}reau}, {Corcione}, {Darmigny}, {de Boer}, {de Teodoro}, {de
  Zeeuw}, {Delle Luche}, {Domingues}, {Dubath}, {Fodor}, {Fr{\'e}zouls},
  {Fries}, {Fustes}, {Fyfe}, {Gallardo}, {Gallegos}, {Gardiol}, {Gebran},
  {Gomboc}, {G{\'o}mez}, {Grux}, {Gueguen}, {Heyrovsky}, {Hoar}, {Iannicola},
  {Isasi Parache}, {Janotto}, {Joliet}, {Jonckheere}, {Keil}, {Kim},
  {Klagyivik}, {Klar}, {Knude}, {Kochukhov}, {Kolka}, {Kos}, {Kutka}, {Lainey},
  {LeBouquin}, {Liu}, {Loreggia}, {Makarov}, {Marseille}, {Martayan},
  {Martinez-Rubi}, {Massart}, {Meynadier}, {Mignot}, {Munari}, {Nguyen},
  {Nordlander}, {Ocvirk}, {O'Flaherty}, {Olias Sanz}, {Ortiz}, {Osorio},
  {Oszkiewicz}, {Ouzounis}, {Palmer}, {Park}, {Pasquato}, {Peltzer}, {Peralta},
  {P{\'e}turaud}, {Pieniluoma}, {Pigozzi}, {Poels}, {Prat}, {Prod'homme},
  {Raison}, {Rebordao}, {Risquez}, {Rocca-Volmerange}, {Rosen}, {Ruiz-Fuertes},
  {Russo}, {Sembay}, {Serraller Vizcaino}, {Short}, {Siebert}, {Silva},
  {Sinachopoulos}, {Slezak}, {Soffel}, {Sosnowska}, {Strai{\v{z}}ys}, {ter
  Linden}, {Terrell}, {Theil}, {Tiede}, {Troisi}, {Tsalmantza}, {Tur},
  {Vaccari}, {Vachier}, {Valles}, {Van Hamme}, {Veltz}, {Virtanen}, {Wallut},
  {Wichmann}, {Wilkinson}, {Ziaeepour}, \& {Zschocke}}]{2016A&A...595A...1G}
{Gaia Collaboration}, {Prusti}, T., {de Bruijne}, J.~H.~J., {et~al.} 2016,
  \aap, 595, A1, \dodoi{10.1051/0004-6361/201629272}

\bibitem[{{Gaia Collaboration} {et~al.}(2023){Gaia Collaboration}, {Vallenari},
  {Brown}, {Prusti}, {de Bruijne}, {Arenou}, {Babusiaux}, {Biermann},
  {Creevey}, {Ducourant}, {Evans}, {Eyer}, {Guerra}, {Hutton}, {Jordi},
  {Klioner}, {Lammers}, {Lindegren}, {Luri}, {Mignard}, {Panem}, {Pourbaix},
  {Randich}, {Sartoretti}, {Soubiran}, {Tanga}, {Walton}, {Bailer-Jones},
  {Bastian}, {Drimmel}, {Jansen}, {Katz}, {Lattanzi}, {van Leeuwen}, {Bakker},
  {Cacciari}, {Casta{\~n}eda}, {De Angeli}, {Fabricius}, {Fouesneau},
  {Fr{\'e}mat}, {Galluccio}, {Guerrier}, {Heiter}, {Masana}, {Messineo},
  {Mowlavi}, {Nicolas}, {Nienartowicz}, {Pailler}, {Panuzzo}, {Riclet}, {Roux},
  {Seabroke}, {Sordo}, {Th{\'e}venin}, {Gracia-Abril}, {Portell}, {Teyssier},
  {Altmann}, {Andrae}, {Audard}, {Bellas-Velidis}, {Benson}, {Berthier},
  {Blomme}, {Burgess}, {Busonero}, {Busso}, {C{\'a}novas}, {Carry}, {Cellino},
  {Cheek}, {Clementini}, {Damerdji}, {Davidson}, {de Teodoro}, {Nu{\~n}ez
  Campos}, {Delchambre}, {Dell'Oro}, {Esquej}, {Fern{\'a}ndez-Hern{\'a}ndez},
  {Fraile}, {Garabato}, {Garc{\'\i}a-Lario}, {Gosset}, {Haigron}, {Halbwachs},
  {Hambly}, {Harrison}, {Hern{\'a}ndez}, {Hestroffer}, {Hodgkin}, {Holl},
  {Jan{\ss}en}, {Jevardat de Fombelle}, {Jordan}, {Krone-Martins}, {Lanzafame},
  {L{\"o}ffler}, {Marchal}, {Marrese}, {Moitinho}, {Muinonen}, {Osborne},
  {Pancino}, {Pauwels}, {Recio-Blanco}, {Reyl{\'e}}, {Riello}, {Rimoldini},
  {Roegiers}, {Rybizki}, {Sarro}, {Siopis}, {Smith}, {Sozzetti}, {Utrilla},
  {van Leeuwen}, {Abbas}, {{\'A}brah{\'a}m}, {Abreu Aramburu}, {Aerts},
  {Aguado}, {Ajaj}, {Aldea-Montero}, {Altavilla}, {{\'A}lvarez}, {Alves},
  {Anders}, {Anderson}, {Anglada Varela}, {Antoja}, {Baines}, {Baker},
  {Balaguer-N{\'u}{\~n}ez}, {Balbinot}, {Balog}, {Barache}, {Barbato},
  {Barros}, {Barstow}, {Bartolom{\'e}}, {Bassilana}, {Bauchet}, {Becciani},
  {Bellazzini}, {Berihuete}, {Bernet}, {Bertone}, {Bianchi}, {Binnenfeld},
  {Blanco-Cuaresma}, {Blazere}, {Boch}, {Bombrun}, {Bossini}, {Bouquillon},
  {Bragaglia}, {Bramante}, {Breedt}, {Bressan}, {Brouillet}, {Brugaletta},
  {Bucciarelli}, {Burlacu}, {Butkevich}, {Buzzi}, {Caffau}, {Cancelliere},
  {Cantat-Gaudin}, {Carballo}, {Carlucci}, {Carnerero}, {Carrasco},
  {Casamiquela}, {Castellani}, {Castro-Ginard}, {Chaoul}, {Charlot}, {Chemin},
  {Chiaramida}, {Chiavassa}, {Chornay}, {Comoretto}, {Contursi}, {Cooper},
  {Cornez}, {Cowell}, {Crifo}, {Cropper}, {Crosta}, {Crowley}, {Dafonte},
  {Dapergolas}, {David}, {David}, {de Laverny}, {De Luise}, {De March}, {De
  Ridder}, {de Souza}, {de Torres}, {del Peloso}, {del Pozo}, {Delbo},
  {Delgado}, {Delisle}, {Demouchy}, {Dharmawardena}, {Di Matteo}, {Diakite},
  {Diener}, {Distefano}, {Dolding}, {Edvardsson}, {Enke}, {Fabre}, {Fabrizio},
  {Faigler}, {Fedorets}, {Fernique}, {Fienga}, {Figueras}, {Fournier},
  {Fouron}, {Fragkoudi}, {Gai}, {Garcia-Gutierrez}, {Garcia-Reinaldos},
  {Garc{\'\i}a-Torres}, {Garofalo}, {Gavel}, {Gavras}, {Gerlach}, {Geyer},
  {Giacobbe}, {Gilmore}, {Girona}, {Giuffrida}, {Gomel}, {Gomez},
  {Gonz{\'a}lez-N{\'u}{\~n}ez}, {Gonz{\'a}lez-Santamar{\'\i}a},
  {Gonz{\'a}lez-Vidal}, {Granvik}, {Guillout}, {Guiraud},
  {Guti{\'e}rrez-S{\'a}nchez}, {Guy}, {Hatzidimitriou}, {Hauser}, {Haywood},
  {Helmer}, {Helmi}, {Sarmiento}, {Hidalgo}, {Hilger}, {H{\l}adczuk}, {Hobbs},
  {Holland}, {Huckle}, {Jardine}, {Jasniewicz}, {Jean-Antoine Piccolo},
  {Jim{\'e}nez-Arranz}, {Jorissen}, {Juaristi Campillo}, {Julbe}, {Karbevska},
  {Kervella}, {Khanna}, {Kontizas}, {Kordopatis}, {Korn}, {K{\'o}sp{\'a}l},
  {Kostrzewa-Rutkowska}, {Kruszy{\'n}ska}, {Kun}, {Laizeau}, {Lambert},
  {Lanza}, {Lasne}, {Le Campion}, {Lebreton}, {Lebzelter}, {Leccia}, {Leclerc},
  {Lecoeur-Taibi}, {Liao}, {Licata}, {Lindstr{\o}m}, {Lister}, {Livanou},
  {Lobel}, {Lorca}, {Loup}, {Madrero Pardo}, {Magdaleno Romeo}, {Managau},
  {Mann}, {Manteiga}, {Marchant}, {Marconi}, {Marcos}, {Marcos Santos},
  {Mar{\'\i}n Pina}, {Marinoni}, {Marocco}, {Marshall}, {Martin Polo},
  {Mart{\'\i}n-Fleitas}, {Marton}, {Mary}, {Masip}, {Massari},
  {Mastrobuono-Battisti}, {Mazeh}, {McMillan}, {Messina}, {Michalik}, {Millar},
  {Mints}, {Molina}, {Molinaro}, {Moln{\'a}r}, {Monari}, {Mongui{\'o}},
  {Montegriffo}, {Montero}, {Mor}, {Mora}, {Morbidelli}, {Morel}, {Morris},
  {Muraveva}, {Murphy}, {Musella}, {Nagy}, {Noval}, {Oca{\~n}a}, {Ogden},
  {Ordenovic}, {Osinde}, {Pagani}, {Pagano}, {Palaversa}, {Palicio},
  {Pallas-Quintela}, {Panahi}, {Payne-Wardenaar}, {Pe{\~n}alosa Esteller},
  {Penttil{\"a}}, {Pichon}, {Piersimoni}, {Pineau}, {Plachy}, {Plum}, {Poggio},
  {Pr{\v{s}}a}, {Pulone}, {Racero}, {Ragaini}, {Rainer}, {Raiteri}, {Rambaux},
  {Ramos}, {Ramos-Lerate}, {Re Fiorentin}, {Regibo}, {Richards}, {Rios Diaz},
  {Ripepi}, {Riva}, {Rix}, {Rixon}, {Robichon}, {Robin}, {Robin}, {Roelens},
  {Rogues}, {Rohrbasser}, {Romero-G{\'o}mez}, {Rowell}, {Royer}, {Ruz Mieres},
  {Rybicki}, {Sadowski}, {S{\'a}ez N{\'u}{\~n}ez}, {Sagrist{\`a} Sell{\'e}s},
  {Sahlmann}, {Salguero}, {Samaras}, {Sanchez Gimenez}, {Sanna},
  {Santove{\~n}a}, {Sarasso}, {Schultheis}, {Sciacca}, {Segol}, {Segovia},
  {S{\'e}gransan}, {Semeux}, {Shahaf}, {Siddiqui}, {Siebert}, {Siltala},
  {Silvelo}, {Slezak}, {Slezak}, {Smart}, {Snaith}, {Solano}, {Solitro},
  {Souami}, {Souchay}, {Spagna}, {Spina}, {Spoto}, {Steele},
  {Steidelm{\"u}ller}, {Stephenson}, {S{\"u}veges}, {Surdej}, {Szabados},
  {Szegedi-Elek}, {Taris}, {Taylor}, {Teixeira}, {Tolomei}, {Tonello}, {Torra},
  {Torra}, {Torralba Elipe}, {Trabucchi}, {Tsounis}, {Turon}, {Ulla}, {Unger},
  {Vaillant}, {van Dillen}, {van Reeven}, {Vanel}, {Vecchiato}, {Viala},
  {Vicente}, {Voutsinas}, {Weiler}, {Wevers}, {Wyrzykowski}, {Yoldas}, {Yvard},
  {Zhao}, {Zorec}, {Zucker}, \& {Zwitter}}]{2023A&A...674A...1G}
{Gaia Collaboration}, {Vallenari}, A., {Brown}, A.~G.~A., {et~al.} 2023, \aap,
  674, A1, \dodoi{10.1051/0004-6361/202243940}

\bibitem[{{Gehrels} {et~al.}(2004){Gehrels}, {Chincarini}, {Giommi}, {Mason},
  {Nousek}, {Wells}, {White}, {Barthelmy}, {Burrows}, {Cominsky}, {Hurley},
  {Marshall}, {M{\'e}sz{\'a}ros}, {Roming}, {Angelini}, {Barbier}, {Belloni},
  {Campana}, {Caraveo}, {Chester}, {Citterio}, {Cline}, {Cropper}, {Cummings},
  {Dean}, {Feigelson}, {Fenimore}, {Frail}, {Fruchter}, {Garmire}, {Gendreau},
  {Ghisellini}, {Greiner}, {Hill}, {Hunsberger}, {Krimm}, {Kulkarni}, {Kumar},
  {Lebrun}, {Lloyd-Ronning}, {Markwardt}, {Mattson}, {Mushotzky}, {Norris},
  {Osborne}, {Paczynski}, {Palmer}, {Park}, {Parsons}, {Paul}, {Rees},
  {Reynolds}, {Rhoads}, {Sasseen}, {Schaefer}, {Short}, {Smale}, {Smith},
  {Stella}, {Tagliaferri}, {Takahashi}, {Tashiro}, {Townsley}, {Tueller},
  {Turner}, {Vietri}, {Voges}, {Ward}, {Willingale}, {Zerbi}, \&
  {Zhang}}]{2004ApJ...611.1005G}
{Gehrels}, N., {Chincarini}, G., {Giommi}, P., {et~al.} 2004, \apj, 611, 1005,
  \dodoi{10.1086/422091}

\bibitem[{{Ghirlanda} \& {Salvaterra}(2022)}]{2022arXiv220606390G}
{Ghirlanda}, G., \& {Salvaterra}, R. 2022, arXiv e-prints, arXiv:2206.06390,
  \dodoi{10.48550/arXiv.2206.06390}

\bibitem[{{Granot} \& {Ramirez-Ruiz}(2010)}]{2010arXiv1012.5101G}
{Granot}, J., \& {Ramirez-Ruiz}, E. 2010, arXiv e-prints, arXiv:1012.5101,
  \dodoi{10.48550/arXiv.1012.5101}

\bibitem[{{Heise}(2003)}]{2003AIPC..662..229H}
{Heise}, J. 2003, in American Institute of Physics Conference Series, Vol. 662,
  Gamma-Ray Burst and Afterglow Astronomy 2001: A Workshop Celebrating the
  First Year of the HETE Mission, ed. G.~R. {Ricker} \& R.~K. {Vanderspek},
  229--236, \dodoi{10.1063/1.1579346}

\bibitem[{{Ho} {et~al.}(2018){Ho}, {Kulkarni}, {Nugent}, {Zhao}, {Rusu},
  {Cenko}, {Ravi}, {Kasliwal}, {Perley}, {Adams}, {Bellm}, {Brady}, {Fremling},
  {Gal-Yam}, {Kann}, {Kaplan}, {Laher}, {Masci}, {Ofek}, {Sollerman}, \&
  {Urban}}]{2018ApJ...854L..13H}
{Ho}, A. Y.~Q., {Kulkarni}, S.~R., {Nugent}, P.~E., {et~al.} 2018, \apjl, 854,
  L13, \dodoi{10.3847/2041-8213/aaaa62}

\bibitem[{{Ho} {et~al.}(2020){Ho}, {Kulkarni}, {Perley}, {Cenko}, {Corsi},
  {Schulze}, {Lunnan}, {Sollerman}, {Gal-Yam}, {Anand}, {Barbarino}, {Bellm},
  {Bruch}, {Burns}, {De}, {Dekany}, {Delacroix}, {Duev}, {Frederiks},
  {Fremling}, {Goldstein}, {Golkhou}, {Graham}, {Hale}, {Kasliwal}, {Kupfer},
  {Laher}, {Martikainen}, {Masci}, {Neill}, {Ridnaia}, {Rusholme}, {Savchenko},
  {Shupe}, {Soumagnac}, {Strotjohann}, {Svinkin}, {Taggart}, {Tartaglia},
  {Yan}, \& {Zolkower}}]{2020ApJ...902...86H}
{Ho}, A. Y.~Q., {Kulkarni}, S.~R., {Perley}, D.~A., {et~al.} 2020, \apj, 902,
  86, \dodoi{10.3847/1538-4357/aba630}

\bibitem[{{Ho} {et~al.}(2022){Ho}, {Perley}, {Yao}, {Svinkin}, {de Ugarte
  Postigo}, {Perley}, {Kann}, {Burns}, {Andreoni}, {Bellm}, {Bissaldi},
  {Bloom}, {Brink}, {Dekany}, {Drake}, {Ag{\"u}{\'\i} Fern{\'a}ndez},
  {Filippenko}, {Frederiks}, {Graham}, {Hristov}, {Kasliwal}, {Kulkarni},
  {Kumar}, {Laher}, {Lysenko}, {Mailyan}, {Malacaria}, {Miller}, {Poolakkil},
  {Riddle}, {Ridnaia}, {Rusholme}, {Savchenko}, {Sollerman}, {Th{\"o}ne},
  {Tsvetkova}, {Ulanov}, \& {von Kienlin}}]{2022ApJ...938...85H}
{Ho}, A. Y.~Q., {Perley}, D.~A., {Yao}, Y., {et~al.} 2022, \apj, 938, 85,
  \dodoi{10.3847/1538-4357/ac8bd0}

\bibitem[{{Hogg} {et~al.}(2002){Hogg}, {Baldry}, {Blanton}, \&
  {Eisenstein}}]{2002astro.ph.10394H}
{Hogg}, D.~W., {Baldry}, I.~K., {Blanton}, M.~R., \& {Eisenstein}, D.~J. 2002,
  arXiv e-prints, astro, \dodoi{10.48550/arXiv.astro-ph/0210394}

\bibitem[{{Hopkins} \& {Beacom}(2006)}]{2006ApJ...651..142H}
{Hopkins}, A.~M., \& {Beacom}, J.~F. 2006, \apj, 651, 142,
  \dodoi{10.1086/506610}

\bibitem[{{Howard} {et~al.}(2019){Howard}, {Corbett}, {Law}, {Ratzloff},
  {Glazier}, {Fors}, {del Ser}, \& {Haislip}}]{2019ApJ...881....9H}
{Howard}, W.~S., {Corbett}, H., {Law}, N.~M., {et~al.} 2019, \apj, 881, 9,
  \dodoi{10.3847/1538-4357/ab2767}

\bibitem[{{Howell}(1989)}]{1989PASP..101..616H}
{Howell}, S.~B. 1989, \pasp, 101, 616, \dodoi{10.1086/132477}

\bibitem[{{Huang} {et~al.}(2002){Huang}, {Dai}, \& {Lu}}]{2002MNRAS.332..735H}
{Huang}, Y.~F., {Dai}, Z.~G., \& {Lu}, T. 2002, \mnras, 332, 735,
  \dodoi{10.1046/j.1365-8711.2002.05334.x}

\bibitem[{{Jakobsson} {et~al.}(2004){Jakobsson}, {Hjorth}, {Fynbo}, {Watson},
  {Pedersen}, {Bj{\"o}rnsson}, \& {Gorosabel}}]{2004ApJ...617L..21J}
{Jakobsson}, P., {Hjorth}, J., {Fynbo}, J.~P.~U., {et~al.} 2004, \apjl, 617,
  L21, \dodoi{10.1086/427089}

\bibitem[{{Kann} {et~al.}(2010){Kann}, {Klose}, {Zhang}, {Malesani}, {Nakar},
  {Pozanenko}, {Wilson}, {Butler}, {Jakobsson}, {Schulze}, {Andreev},
  {Antonelli}, {Bikmaev}, {Biryukov}, {B{\"o}ttcher}, {Burenin}, {Castro
  Cer{\'o}n}, {Castro-Tirado}, {Chincarini}, {Cobb}, {Covino}, {D'Avanzo},
  {D'Elia}, {Della Valle}, {de Ugarte Postigo}, {Efimov}, {Ferrero}, {Fugazza},
  {Fynbo}, {G{\r{a}}lfalk}, {Grundahl}, {Gorosabel}, {Gupta}, {Guziy},
  {Hafizov}, {Hjorth}, {Holhjem}, {Ibrahimov}, {Im}, {Israel}, {Je{\'l}inek},
  {Jensen}, {Karimov}, {Khamitov}, {Kizilo{\v{g}}lu}, {Klunko}, {Kub{\'a}nek},
  {Kutyrev}, {Laursen}, {Levan}, {Mannucci}, {Martin}, {Mescheryakov},
  {Mirabal}, {Norris}, {Ovaldsen}, {Paraficz}, {Pavlenko}, {Piranomonte},
  {Rossi}, {Rumyantsev}, {Salinas}, {Sergeev}, {Sharapov}, {Sollerman},
  {Stecklum}, {Stella}, {Tagliaferri}, {Tanvir}, {Telting}, {Testa}, {Updike},
  {Volnova}, {Watson}, {Wiersema}, \& {Xu}}]{2010ApJ...720.1513K}
{Kann}, D.~A., {Klose}, S., {Zhang}, B., {et~al.} 2010, \apj, 720, 1513,
  \dodoi{10.1088/0004-637X/720/2/1513}

\bibitem[{{Karpov} {et~al.}(2017){Karpov}, {Beskin}, {Biryukov}, {Bondar},
  {Ivanov}, {Katkova}, {Orekhova}, {Perkov}, \& {Sasyuk}}]{2017IAUS..324...85K}
{Karpov}, S., {Beskin}, G., {Biryukov}, A., {et~al.} 2017, in New Frontiers in
  Black Hole Astrophysics, ed. A.~{Gomboc}, Vol. 324, 85--86,
  \dodoi{10.1017/S1743921317001259}

\bibitem[{{Kistler} {et~al.}(2008){Kistler}, {Y{\"u}ksel}, {Beacom}, \&
  {Stanek}}]{2008ApJ...673L.119K}
{Kistler}, M.~D., {Y{\"u}ksel}, H., {Beacom}, J.~F., \& {Stanek}, K.~Z. 2008,
  \apjl, 673, L119, \dodoi{10.1086/527671}

\bibitem[{{Klotz} {et~al.}(2009){Klotz}, {Bo{\"e}r}, {Atteia}, \&
  {Gendre}}]{2009AJ....137.4100K}
{Klotz}, A., {Bo{\"e}r}, M., {Atteia}, J.~L., \& {Gendre}, B. 2009, \aj, 137,
  4100, \dodoi{10.1088/0004-6256/137/5/4100}

\bibitem[{{Kopa{\v{c}}} {et~al.}(2013){Kopa{\v{c}}}, {Kobayashi}, {Gomboc},
  {Japelj}, {Mundell}, {Guidorzi}, {Melandri}, {Bersier}, {Cano}, {Smith},
  {Steele}, \& {Virgili}}]{2013ApJ...772...73K}
{Kopa{\v{c}}}, D., {Kobayashi}, S., {Gomboc}, A., {et~al.} 2013, \apj, 772, 73,
  \dodoi{10.1088/0004-637X/772/1/73}

\bibitem[{{Kumar} \& {Zhang}(2015)}]{2015PhR...561....1K}
{Kumar}, P., \& {Zhang}, B. 2015, \physrep, 561, 1,
  \dodoi{10.1016/j.physrep.2014.09.008}

\bibitem[{{Lamb} \& {Reichart}(2000)}]{2000ApJ...536....1L}
{Lamb}, D.~Q., \& {Reichart}, D.~E. 2000, \apj, 536, 1, \dodoi{10.1086/308918}

\bibitem[{{Langer} \& {Norman}(2006)}]{2006ApJ...638L..63L}
{Langer}, N., \& {Norman}, C.~A. 2006, \apjl, 638, L63, \dodoi{10.1086/500363}

\bibitem[{{Law} {et~al.}(2015){Law}, {Fors}, {Ratzloff}, {Wulfken},
  {Kavanaugh}, {Sitar}, {Pruett}, {Birchard}, {Barlow}, {Cannon}, {Cenko},
  {Dunlap}, {Kraus}, \& {Maccarone}}]{2015PASP..127..234L}
{Law}, N.~M., {Fors}, O., {Ratzloff}, J., {et~al.} 2015, \pasp, 127, 234,
  \dodoi{10.1086/680521}

\bibitem[{{Law} {et~al.}(2022){Law}, {Corbett}, {Galliher}, {Gonzalez},
  {Vasquez}, {Walters}, {Machia}, {Ratzloff}, {Ackley}, {Bizon}, {Clemens},
  {Cox}, {Eikenberry}, {Howard}, {Glazier}, {Mann}, {Quimby}, {Reichart}, \&
  {Trilling}}]{2022PASP..134c5003L}
{Law}, N.~M., {Corbett}, H., {Galliher}, N.~W., {et~al.} 2022, \pasp, 134,
  035003, \dodoi{10.1088/1538-3873/ac4811}

\bibitem[{{Li} {et~al.}(2012){Li}, {Liang}, {Tang}, {Chen}, {Xi}, {L{\"u}},
  {Gao}, {Zhang}, {Zhang}, {Yi}, {Lu}, {L{\"u}}, \&
  {Wei}}]{2012ApJ...758...27L}
{Li}, L., {Liang}, E.-W., {Tang}, Q.-W., {et~al.} 2012, \apj, 758, 27,
  \dodoi{10.1088/0004-637X/758/1/27}

\bibitem[{{Lipunov} {et~al.}(2007){Lipunov}, {Kornilov}, {Krylov}, {Tyurina},
  {Belinskii}, {Gorbovskoi}, {Kuvshinov}, {Gritsyk}, {Antipov}, {Borisov},
  {Sankovich}, {Vladimirov}, {Vybornov}, \& {Kuznetsov}}]{2007ARep...51.1004L}
{Lipunov}, V.~M., {Kornilov}, V.~G., {Krylov}, A.~V., {et~al.} 2007, Astronomy
  Reports, 51, 1004, \dodoi{10.1134/S1063772907120050}

\bibitem[{{Meegan} {et~al.}(2009){Meegan}, {Lichti}, {Bhat}, {Bissaldi},
  {Briggs}, {Connaughton}, {Diehl}, {Fishman}, {Greiner}, {Hoover}, {van der
  Horst}, {von Kienlin}, {Kippen}, {Kouveliotou}, {McBreen}, {Paciesas},
  {Preece}, {Steinle}, {Wallace}, {Wilson}, \&
  {Wilson-Hodge}}]{2009ApJ...702..791M}
{Meegan}, C., {Lichti}, G., {Bhat}, P.~N., {et~al.} 2009, \apj, 702, 791,
  \dodoi{10.1088/0004-637X/702/1/791}

\bibitem[{{M{\'e}sz{\'a}ros} \& {Rees}(1997)}]{1997ApJ...482L..29M}
{M{\'e}sz{\'a}ros}, P., \& {Rees}, M.~J. 1997, \apjl, 482, L29,
  \dodoi{10.1086/310692}

\bibitem[{{Nemiroff}(2003)}]{2003AJ....125.2740N}
{Nemiroff}, R.~J. 2003, \aj, 125, 2740, \dodoi{10.1086/374571}

\bibitem[{{Nir} {et~al.}(2021){Nir}, {Ofek}, {Ben-Ami}, {Segev}, {Polishook},
  \& {Manulis}}]{2021MNRAS.505.2477N}
{Nir}, G., {Ofek}, E.~O., {Ben-Ami}, S., {et~al.} 2021, \mnras, 505, 2477,
  \dodoi{10.1093/mnras/stab1437}

\bibitem[{{Oates}(2023)}]{2023Univ....9..113O}
{Oates}, S. 2023, Universe, 9, 113, \dodoi{10.3390/universe9030113}

\bibitem[{{Ofek} \& {Ben-Ami}(2020)}]{2020PASP..132l5004O}
{Ofek}, E.~O., \& {Ben-Ami}, S. 2020, \pasp, 132, 125004,
  \dodoi{10.1088/1538-3873/abc14c}

\bibitem[{{Ofek} {et~al.}(2023){Ofek}, {Ben-Ami}, {Polishook}, {Segre},
  {Blumenzweig}, {Strotjohann}, {Yaron}, {Shani}, {Nachshon}, {Shvartzvald},
  {Hershko}, {Engel}, {Segre}, {Segev}, {Zimmerman}, {Nir}, {Judkovsky},
  {Gal-Yam}, {Zackay}, {Waxman}, {Kushnir}, {Chen}, {Azaria}, {Manulis},
  {Diner}, {Vandeventer}, {Franckowiak}, {Weimann}, {Borowska}, {Garrappa},
  {Zenin}, {Fallah Ramazani}, {Konno}, {K{\"u}sters}, {Sadeh}, {Parsons},
  {Berge}, {Kowalski}, {Ohm}, {Arcavi}, \& {Bruch}}]{2023PASP..135f5001O}
{Ofek}, E.~O., {Ben-Ami}, S., {Polishook}, D., {et~al.} 2023, \pasp, 135,
  065001, \dodoi{10.1088/1538-3873/acd8f0}

\bibitem[{{Oganesyan} {et~al.}(2019){Oganesyan}, {Nava}, {Ghirlanda},
  {Melandri}, \& {Celotti}}]{2019A&A...628A..59O}
{Oganesyan}, G., {Nava}, L., {Ghirlanda}, G., {Melandri}, A., \& {Celotti}, A.
  2019, \aap, 628, A59, \dodoi{10.1051/0004-6361/201935766}

\bibitem[{{Oganesyan} {et~al.}(2023){Oganesyan}, {Karpov}, {Salafia},
  {Jel{\'\i}nek}, {Beskin}, {Ronchini}, {Banerjee}, {Branchesi},
  {{\v{S}}trobl}, {Pol{\'a}{\v{s}}ek}, {Hudec}, {Ivanov}, {Katkova}, {Perkov},
  {Biryukov}, {Lyapsina}, {Sasyuk}, {Ma{\v{s}}ek}, {Jane{\v{c}}ek}, {Ebr},
  {Jury{\v{s}}ek}, {Cunniffe}, \& {Prouza}}]{2023NatAs...7..843O}
{Oganesyan}, G., {Karpov}, S., {Salafia}, O.~S., {et~al.} 2023, Nature
  Astronomy, 7, 843, \dodoi{10.1038/s41550-023-01972-4}

\bibitem[{{Panaitescu} \& {Vestrand}(2008)}]{2008MNRAS.387..497P}
{Panaitescu}, A., \& {Vestrand}, W.~T. 2008, \mnras, 387, 497,
  \dodoi{10.1111/j.1365-2966.2008.13231.x}

\bibitem[{{Panaitescu} \& {Vestrand}(2011)}]{2011MNRAS.414.3537P}
---. 2011, \mnras, 414, 3537, \dodoi{10.1111/j.1365-2966.2011.18653.x}

\bibitem[{{Pescalli} {et~al.}(2015){Pescalli}, {Ghirlanda}, {Salafia},
  {Ghisellini}, {Nappo}, \& {Salvaterra}}]{2015MNRAS.447.1911P}
{Pescalli}, A., {Ghirlanda}, G., {Salafia}, O.~S., {et~al.} 2015, \mnras, 447,
  1911, \dodoi{10.1093/mnras/stu2482}

\bibitem[{{Piran}(2004)}]{2004RvMP...76.1143P}
{Piran}, T. 2004, Reviews of Modern Physics, 76, 1143,
  \dodoi{10.1103/RevModPhys.76.1143}

\bibitem[{{Planck Collaboration} {et~al.}(2020){Planck Collaboration},
  {Aghanim}, {Akrami}, {Ashdown}, {Aumont}, {Baccigalupi}, {Ballardini},
  {Banday}, {Barreiro}, {Bartolo}, {Basak}, {Battye}, {Benabed}, {Bernard},
  {Bersanelli}, {Bielewicz}, {Bock}, {Bond}, {Borrill}, {Bouchet}, {Boulanger},
  {Bucher}, {Burigana}, {Butler}, {Calabrese}, {Cardoso}, {Carron},
  {Challinor}, {Chiang}, {Chluba}, {Colombo}, {Combet}, {Contreras}, {Crill},
  {Cuttaia}, {de Bernardis}, {de Zotti}, {Delabrouille}, {Delouis}, {Di
  Valentino}, {Diego}, {Dor{\'e}}, {Douspis}, {Ducout}, {Dupac}, {Dusini},
  {Efstathiou}, {Elsner}, {En{\ss}lin}, {Eriksen}, {Fantaye}, {Farhang},
  {Fergusson}, {Fernandez-Cobos}, {Finelli}, {Forastieri}, {Frailis},
  {Fraisse}, {Franceschi}, {Frolov}, {Galeotta}, {Galli}, {Ganga},
  {G{\'e}nova-Santos}, {Gerbino}, {Ghosh}, {Gonz{\'a}lez-Nuevo}, {G{\'o}rski},
  {Gratton}, {Gruppuso}, {Gudmundsson}, {Hamann}, {Handley}, {Hansen},
  {Herranz}, {Hildebrandt}, {Hivon}, {Huang}, {Jaffe}, {Jones}, {Karakci},
  {Keih{\"a}nen}, {Keskitalo}, {Kiiveri}, {Kim}, {Kisner}, {Knox},
  {Krachmalnicoff}, {Kunz}, {Kurki-Suonio}, {Lagache}, {Lamarre}, {Lasenby},
  {Lattanzi}, {Lawrence}, {Le Jeune}, {Lemos}, {Lesgourgues}, {Levrier},
  {Lewis}, {Liguori}, {Lilje}, {Lilley}, {Lindholm}, {L{\'o}pez-Caniego},
  {Lubin}, {Ma}, {Mac{\'\i}as-P{\'e}rez}, {Maggio}, {Maino}, {Mandolesi},
  {Mangilli}, {Marcos-Caballero}, {Maris}, {Martin}, {Martinelli},
  {Mart{\'\i}nez-Gonz{\'a}lez}, {Matarrese}, {Mauri}, {McEwen}, {Meinhold},
  {Melchiorri}, {Mennella}, {Migliaccio}, {Millea}, {Mitra},
  {Miville-Desch{\^e}nes}, {Molinari}, {Montier}, {Morgante}, {Moss}, {Natoli},
  {N{\o}rgaard-Nielsen}, {Pagano}, {Paoletti}, {Partridge}, {Patanchon},
  {Peiris}, {Perrotta}, {Pettorino}, {Piacentini}, {Polastri}, {Polenta},
  {Puget}, {Rachen}, {Reinecke}, {Remazeilles}, {Renzi}, {Rocha}, {Rosset},
  {Roudier}, {Rubi{\~n}o-Mart{\'\i}n}, {Ruiz-Granados}, {Salvati}, {Sandri},
  {Savelainen}, {Scott}, {Shellard}, {Sirignano}, {Sirri}, {Spencer},
  {Sunyaev}, {Suur-Uski}, {Tauber}, {Tavagnacco}, {Tenti}, {Toffolatti},
  {Tomasi}, {Trombetti}, {Valenziano}, {Valiviita}, {Van Tent}, {Vibert},
  {Vielva}, {Villa}, {Vittorio}, {Wandelt}, {Wehus}, {White}, {White},
  {Zacchei}, \& {Zonca}}]{2020A&A...641A...6P}
{Planck Collaboration}, {Aghanim}, N., {Akrami}, Y., {et~al.} 2020, \aap, 641,
  A6, \dodoi{10.1051/0004-6361/201833910}

\bibitem[{{Qin} {et~al.}(2010){Qin}, {Liang}, {Lu}, {Wei}, \&
  {Zhang}}]{2010MNRAS.406..558Q}
{Qin}, S.-F., {Liang}, E.-W., {Lu}, R.-J., {Wei}, J.-Y., \& {Zhang}, S.-N.
  2010, \mnras, 406, 558, \dodoi{10.1111/j.1365-2966.2010.16691.x}

\bibitem[{{Racusin} {et~al.}(2008){Racusin}, {Karpov}, {Sokolowski}, {Granot},
  {Wu}, {Pal'Shin}, {Covino}, {van der Horst}, {Oates}, {Schady}, {Smith},
  {Cummings}, {Starling}, {Piotrowski}, {Zhang}, {Evans}, {Holland}, {Malek},
  {Page}, {Vetere}, {Margutti}, {Guidorzi}, {Kamble}, {Curran}, {Beardmore},
  {Kouveliotou}, {Mankiewicz}, {Melandri}, {O'Brien}, {Page}, {Piran},
  {Tanvir}, {Wrochna}, {Aptekar}, {Barthelmy}, {Bartolini}, {Beskin}, {Bondar},
  {Bremer}, {Campana}, {Castro-Tirado}, {Cucchiara}, {Cwiok}, {D'Avanzo},
  {D'Elia}, {Della Valle}, {de Ugarte Postigo}, {Dominik}, {Falcone}, {Fiore},
  {Fox}, {Frederiks}, {Fruchter}, {Fugazza}, {Garrett}, {Gehrels},
  {Golenetskii}, {Gomboc}, {Gorosabel}, {Greco}, {Guarnieri}, {Immler},
  {Jelinek}, {Kasprowicz}, {La Parola}, {Levan}, {Mangano}, {Mazets},
  {Molinari}, {Moretti}, {Nawrocki}, {Oleynik}, {Osborne}, {Pagani}, {Pandey},
  {Paragi}, {Perri}, {Piccioni}, {Ramirez-Ruiz}, {Roming}, {Steele}, {Strom},
  {Testa}, {Tosti}, {Ulanov}, {Wiersema}, {Wijers}, {Winters}, {Zarnecki},
  {Zerbi}, {M{\'e}sz{\'a}ros}, {Chincarini}, \&
  {Burrows}}]{2008Natur.455..183R}
{Racusin}, J.~L., {Karpov}, S.~V., {Sokolowski}, M., {et~al.} 2008, \nat, 455,
  183, \dodoi{10.1038/nature07270}

\bibitem[{{Rees} \& {Meszaros}(1994)}]{1994ApJ...430L..93R}
{Rees}, M.~J., \& {Meszaros}, P. 1994, \apjl, 430, L93, \dodoi{10.1086/187446}

\bibitem[{{Rudolph} {et~al.}(2022){Rudolph}, {Bo{\v{s}}njak}, {Palladino},
  {Sadeh}, \& {Winter}}]{2022MNRAS.511.5823R}
{Rudolph}, A., {Bo{\v{s}}njak}, {\v{Z}}., {Palladino}, A., {Sadeh}, I., \&
  {Winter}, W. 2022, \mnras, 511, 5823, \dodoi{10.1093/mnras/stac433}

\bibitem[{{Rudolph} {et~al.}(2023){Rudolph}, {Petropoulou}, {Bo{\v{s}}njak}, \&
  {Winter}}]{2023ApJ...950...28R}
{Rudolph}, A., {Petropoulou}, M., {Bo{\v{s}}njak}, {\v{Z}}., \& {Winter}, W.
  2023, \apj, 950, 28, \dodoi{10.3847/1538-4357/acc861}

\bibitem[{{Rykoff} {et~al.}(2005){Rykoff}, {Aharonian}, {Akerlof}, {Alatalo},
  {Ashley}, {G{\"u}ver}, {Horns}, {Kehoe}, {Kizilo{\v{g}}lu}, {McKay},
  {{\"O}zel}, {Phillips}, {Quimby}, {Schaefer}, {Smith}, {Swan}, {Vestrand},
  {Wheeler}, {Wren}, \& {Yost}}]{2005ApJ...631.1032R}
{Rykoff}, E.~S., {Aharonian}, F., {Akerlof}, C.~W., {et~al.} 2005, \apj, 631,
  1032, \dodoi{10.1086/432832}

\bibitem[{{Sadeh}(2020)}]{2020ApJ...894L..25S}
{Sadeh}, I. 2020, \apjl, 894, L25, \dodoi{10.3847/2041-8213/ab8b5f}

\bibitem[{{Sakamoto} {et~al.}(2005){Sakamoto}, {Lamb}, {Kawai}, {Yoshida},
  {Graziani}, {Fenimore}, {Donaghy}, {Matsuoka}, {Suzuki}, {Ricker}, {Atteia},
  {Shirasaki}, {Tamagawa}, {Torii}, {Galassi}, {Doty}, {Vanderspek}, {Crew},
  {Villasenor}, {Butler}, {Prigozhin}, {Jernigan}, {Barraud}, {Boer},
  {Dezalay}, {Olive}, {Hurley}, {Levine}, {Monnelly}, {Martel}, {Morgan},
  {Woosley}, {Cline}, {Braga}, {Manchanda}, {Pizzichini}, {Takagishi}, \&
  {Yamauchi}}]{2005ApJ...629..311S}
{Sakamoto}, T., {Lamb}, D.~Q., {Kawai}, N., {et~al.} 2005, \apj, 629, 311,
  \dodoi{10.1086/431235}

\bibitem[{{Salvaterra} {et~al.}(2012){Salvaterra}, {Campana}, {Vergani},
  {Covino}, {D'Avanzo}, {Fugazza}, {Ghirlanda}, {Ghisellini}, {Melandri},
  {Nava}, {Sbarufatti}, {Flores}, {Piranomonte}, \&
  {Tagliaferri}}]{2012ApJ...749...68S}
{Salvaterra}, R., {Campana}, S., {Vergani}, S.~D., {et~al.} 2012, \apj, 749,
  68, \dodoi{10.1088/0004-637X/749/1/68}

\bibitem[{{Senno} {et~al.}(2016){Senno}, {Murase}, \&
  {M{\'e}sz{\'a}ros}}]{2016PhRvD..93h3003S}
{Senno}, N., {Murase}, K., \& {M{\'e}sz{\'a}ros}, P. 2016, \prd, 93, 083003,
  \dodoi{10.1103/PhysRevD.93.083003}

\bibitem[{{Shibata} \& {Magara}(2011)}]{2011LRSP....8....6S}
{Shibata}, K., \& {Magara}, T. 2011, Living Reviews in Solar Physics, 8, 6,
  \dodoi{10.12942/lrsp-2011-6}

\bibitem[{{Shvartzvald} {et~al.}(2023){Shvartzvald}, {Waxman}, {Gal-Yam},
  {Ofek}, {Ben-Ami}, {Berge}, {Kowalski}, {B{\"u}hler}, {Worm}, {Rhoads},
  {Arcavi}, {Maoz}, {Polishook}, {Stone}, {Trakhtenbrot}, {Ackermann},
  {Aharonson}, {Birnholtz}, {Chelouche}, {Guetta}, {Hallakoun}, {Horesh},
  {Kushnir}, {Mazeh}, {Nordin}, {Ofir}, {Ohm}, {Parsons}, {Pe'er}, {Perets},
  {Perdelwitz}, {Poznanski}, {Sadeh}, {Sagiv}, {Shahaf}, {Soumagnac}, {Tal-Or},
  {Van Santen}, {Zackay}, {Guttman}, {Rekhi}, {Townsend}, {Weinstein}, \&
  {Wold}}]{2023arXiv230414482S}
{Shvartzvald}, Y., {Waxman}, E., {Gal-Yam}, A., {et~al.} 2023, arXiv e-prints,
  arXiv:2304.14482, \dodoi{10.48550/arXiv.2304.14482}

\bibitem[{{Sokolowski} {et~al.}(2009){Sokolowski}, {Cwiok}, {Dominik},
  {Juchniewicz}, {Kasprowicz}, {Majcher}, {Majczyna}, {Malek}, {Mankiewicz},
  {Nawrocki}, {Pietrzak}, {Piotrowski}, {Rybka}, {Uzycki}, {Wawrzaszek},
  {Wrochna}, {Zaremba}, \& {{\.Z}arnecki}}]{2009AIPC.1133..306S}
{Sokolowski}, M., {Cwiok}, M., {Dominik}, W., {et~al.} 2009, in American
  Institute of Physics Conference Series, Vol. 1133, Gamma-ray Burst: Sixth
  Huntsville Symposium, ed. C.~{Meegan}, C.~{Kouveliotou}, \& N.~{Gehrels},
  306--311, \dodoi{10.1063/1.3155907}

\bibitem[{{Stalder} {et~al.}(2017){Stalder}, {Tonry}, {Smartt}, {Coughlin},
  {Chambers}, {Stubbs}, {Chen}, {Kankare}, {Smith}, {Denneau}, {Sherstyuk},
  {Heinze}, {Weiland}, {Rest}, {Young}, {Huber}, {Flewelling}, {Lowe},
  {Magnier}, {Schultz}, {Waters}, {Wainscoat}, {Willman}, {Wright}, {Chu},
  {Sanders}, {Inserra}, {Maguire}, \& {Kotak}}]{2017ApJ...850..149S}
{Stalder}, B., {Tonry}, J., {Smartt}, S.~J., {et~al.} 2017, \apj, 850, 149,
  \dodoi{10.3847/1538-4357/aa95c1}

\bibitem[{{Steele} {et~al.}(2004){Steele}, {Smith}, {Rees}, {Baker}, {Bates},
  {Bode}, {Bowman}, {Carter}, {Etherton}, {Ford}, {Fraser}, {Gomboc}, {Lett},
  {Mansfield}, {Marchant}, {Medrano-Cerda}, {Mottram}, {Raback}, {Scott},
  {Tomlinson}, \& {Zamanov}}]{2004SPIE.5489..679S}
{Steele}, I.~A., {Smith}, R.~J., {Rees}, P.~C., {et~al.} 2004, in Society of
  Photo-Optical Instrumentation Engineers (SPIE) Conference Series, Vol. 5489,
  Ground-based Telescopes, ed. J.~{Oschmann}, Jacobus~M., 679--692,
  \dodoi{10.1117/12.551456}

\bibitem[{Thompson(1994)}]{10.1093/mnras/270.3.480}
Thompson, C. 1994, Monthly Notices of the Royal Astronomical Society, 270, 480,
  \dodoi{10.1093/mnras/270.3.480}

\bibitem[{{Troja} {et~al.}(2017){Troja}, {Lipunov}, {Mundell}, {Butler},
  {Watson}, {Kobayashi}, {Cenko}, {Marshall}, {Ricci}, {Fruchter}, {Wieringa},
  {Gorbovskoy}, {Kornilov}, {Kutyrev}, {Lee}, {Toy}, {Tyurina}, {Budnev},
  {Buckley}, {Gonz{\'a}lez}, {Gress}, {Horesh}, {Panasyuk}, {Prochaska},
  {Ramirez-Ruiz}, {Rebolo Lopez}, {Richer}, {Roman-Zuniga}, {Serra-Ricart},
  {Yurkov}, \& {Gehrels}}]{2017Natur.547..425T}
{Troja}, E., {Lipunov}, V.~M., {Mundell}, C.~G., {et~al.} 2017, \nat, 547, 425,
  \dodoi{10.1038/nature23289}

\bibitem[{{van Dokkum}(2001)}]{2001PASP..113.1420V}
{van Dokkum}, P.~G. 2001, \pasp, 113, 1420, \dodoi{10.1086/323894}

\bibitem[{{van Roestel} {et~al.}(2019){van Roestel}, {Groot}, {Kupfer},
  {Verbeek}, {van Velzen}, {Bours}, {Nugent}, {Prince}, {Levitan}, {Nissanke},
  {Kulkarni}, \& {Laher}}]{2019MNRAS.484.4507V}
{van Roestel}, J., {Groot}, P.~J., {Kupfer}, T., {et~al.} 2019, \mnras, 484,
  4507, \dodoi{10.1093/mnras/stz241}

\bibitem[{{Vestrand} {et~al.}(2014){Vestrand}, {Wren}, {Panaitescu}, {Wozniak},
  {Davis}, {Palmer}, {Vianello}, {Omodei}, {Xiong}, {Briggs}, {Elphick},
  {Paciesas}, \& {Rosing}}]{2014Sci...343...38V}
{Vestrand}, W.~T., {Wren}, J.~A., {Panaitescu}, A., {et~al.} 2014, Science,
  343, 38, \dodoi{10.1126/science.1242316}

\bibitem[{{Wanderman} \& {Piran}(2010)}]{2010MNRAS.406.1944W}
{Wanderman}, D., \& {Piran}, T. 2010, \mnras, 406, 1944,
  \dodoi{10.1111/j.1365-2966.2010.16787.x}

\bibitem[{{Wang} {et~al.}(2013){Wang}, {Liang}, {Li}, {Lu}, {Wei}, \&
  {Zhang}}]{2013ApJ...774..132W}
{Wang}, X.-G., {Liang}, E.-W., {Li}, L., {et~al.} 2013, \apj, 774, 132,
  \dodoi{10.1088/0004-637X/774/2/132}

\bibitem[{{Waxman} \& {Katz}(2017)}]{2017hsn..book..967W}
{Waxman}, E., \& {Katz}, B. 2017, in Handbook of Supernovae, ed. A.~W.
  {Alsabti} \& P.~{Murdin}, 967, \dodoi{10.1007/978-3-319-21846-5_33}

\bibitem[{{Wei} {et~al.}(2016){Wei}, {Cordier}, {Antier}, {Antilogus},
  {Atteia}, {Bajat}, {Basa}, {Beckmann}, {Bernardini}, {Boissier}, {Bouchet},
  {Burwitz}, {Claret}, {Dai}, {Daigne}, {Deng}, {Dornic}, {Feng}, {Foglizzo},
  {Gao}, {Gehrels}, {Godet}, {Goldwurm}, {Gonzalez}, {Gosset}, {G{\"o}tz},
  {Gouiffes}, {Grise}, {Gros}, {Guilet}, {Han}, {Huang}, {Huang}, {Jouret},
  {Klotz}, {La Marle}, {Lachaud}, {Le Floch}, {Lee}, {Leroy}, {Li}, {Li}, {Li},
  {Liang}, {Lyu}, {Mercier}, {Migliori}, {Mochkovitch}, {O'Brien}, {Osborne},
  {Paul}, {Perinati}, {Petitjean}, {Piron}, {Qiu}, {Rau}, {Rodriguez},
  {Schanne}, {Tanvir}, {Vangioni}, {Vergani}, {Wang}, {Wang}, {Wang}, {Wang},
  {Watson}, {Webb}, {Wei}, {Willingale}, {Wu}, {Wu}, {Xin}, {Xu}, {Yu}, {Yu},
  {Yu}, {Zhang}, {Zhang}, {Zhang}, \& {Zhou}}]{2016arXiv161006892W}
{Wei}, J., {Cordier}, B., {Antier}, S., {et~al.} 2016, arXiv e-prints,
  arXiv:1610.06892, \dodoi{10.48550/arXiv.1610.06892}

\bibitem[{{Woosley}(1993)}]{1993ApJ...405..273W}
{Woosley}, S.~E. 1993, \apj, 405, 273, \dodoi{10.1086/172359}

\bibitem[{{Woosley} \& {Bloom}(2006)}]{2006ARA&A..44..507W}
{Woosley}, S.~E., \& {Bloom}, J.~S. 2006, \araa, 44, 507,
  \dodoi{10.1146/annurev.astro.43.072103.150558}

\bibitem[{{Xin} {et~al.}(2023){Xin}, {Han}, {Li}, {Zhang}, {Wang}, {Turpin},
  {Yang}, {Qiu}, {Liang}, {Dai}, {Cai}, {Lu}, {Wang}, {Huang}, {Wang}, {Wu},
  {Gao}, {Ren}, {Zhang}, {Yang}, {Deng}, \& {Wei}}]{2023NatAs...7..724X}
{Xin}, L., {Han}, X., {Li}, H., {et~al.} 2023, Nature Astronomy, 7, 724,
  \dodoi{10.1038/s41550-023-01930-0}

\bibitem[{{Yost} {et~al.}(2007{\natexlab{a}}){Yost}, {Swan}, {Rykoff},
  {Aharonian}, {Akerlof}, {Alday}, {Ashley}, {Barthelmy}, {Burrows}, {Depoy},
  {Dufour}, {Eastman}, {Forgey}, {Gehrels}, {G{\"o}{\v{g}}{\"u}{\c{s}}},
  {G{\"u}ver}, {Halpern}, {Hardin}, {Horns}, {Kiziloglu}, {Krimm}, {Lepine},
  {Liang}, {Marshall}, {McKay}, {Mineo}, {Mirabal}, {{\"O}zel}, {Phillips},
  {Prieto}, {Quimby}, {Romano}, {Rowell}, {Rujopakarn}, {Schaefer},
  {Silverman}, {Siverd}, {Skinner}, {Smith}, {Smith}, {Tonnesen}, {Troja},
  {Vestrand}, {Wheeler}, {Wren}, {Yuan}, \& {Zhang}}]{2007ApJ...657..925Y}
{Yost}, S.~A., {Swan}, H.~F., {Rykoff}, E.~S., {et~al.} 2007{\natexlab{a}},
  \apj, 657, 925, \dodoi{10.1086/510896}

\bibitem[{{Yost} {et~al.}(2007{\natexlab{b}}){Yost}, {Aharonian}, {Akerlof},
  {Ashley}, {Barthelmy}, {Gehrels}, {G{\"o}{\v{g}}{\"u}{\c{s}}}, {G{\"u}ver},
  {Horns}, {K{\i}z{\i}lo{\v{g}}lu}, {Krimm}, {McKay}, {{\"O}zel}, {Phillips},
  {Quimby}, {Rowell}, {Rujopakarn}, {Rykoff}, {Schaefer}, {Smith}, {Swan},
  {Vestrand}, {Wheeler}, {Wren}, \& {Yuan}}]{2007ApJ...669.1107Y}
{Yost}, S.~A., {Aharonian}, F., {Akerlof}, C.~W., {et~al.} 2007{\natexlab{b}},
  \apj, 669, 1107, \dodoi{10.1086/521668}

\bibitem[{{Zhang} {et~al.}(2018){Zhang}, {Zhang}, {Castro-Tirado}, {Dai},
  {Tam}, {Wang}, {Hu}, {Karpov}, {Pozanenko}, {Zhang}, {Mazaeva}, {Minaev},
  {Volnova}, {Oates}, {Gao}, {Wu}, {Shao}, {Tang}, {Beskin}, {Biryukov},
  {Bondar}, {Ivanov}, {Katkova}, {Orekhova}, {Perkov}, {Sasyuk}, {Mankiewicz},
  {{\.Z}arnecki}, {Cwiek}, {Opiela}, {Zadro{\.Z}ny}, {Aptekar}, {Frederiks},
  {Svinkin}, {Kusakin}, {Inasaridze}, {Burhonov}, {Rumyantsev}, {Klunko},
  {Moskvitin}, {Fatkhullin}, {Sokolov}, {Valeev}, {Jeong}, {Park},
  {Caballero-Garc{\'\i}a}, {Cunniffe}, {Tello}, {Ferrero}, {Pandey},
  {Jel{\'\i}nek}, {Peng}, {S{\'a}nchez-Ram{\'\i}rez}, \&
  {Castell{\'o}n}}]{2018NatAs...2...69Z}
{Zhang}, B.~B., {Zhang}, B., {Castro-Tirado}, A.~J., {et~al.} 2018, Nature
  Astronomy, 2, 69, \dodoi{10.1038/s41550-017-0309-8}

\end{thebibliography}
\bibliographystyle{aasjournal}

\end{document}